\documentclass[a4paper,12pt]{article}
\pdfoutput=1

\usepackage{jheppub}
\usepackage{bm,latexsym,amsmath,amssymb,amsfonts,mathtools,mathrsfs}

\usepackage{graphicx}
\graphicspath{{./Figures/}}
\usepackage{wrapfig}

\usepackage[hang]{subfigure}
\usepackage{color}

\usepackage[export]{adjustbox}

\usepackage{todonotes}
\definecolor{palatinate}{RGB}{128, 49, 123}

\definecolor{nicered}{rgb}{0.7,0.1,0.1}
\definecolor{nicegreen}{rgb}{0.1,0.5,0.1}

\newcommand{\be}{\begin{equation}}
\newcommand{\ee}{\end{equation}}
\newcommand{\beal}{\begin{aligned}}
\newcommand{\eeal}{\end{aligned}}
\newcommand\bea{\begin{eqnarray}}
\newcommand\eea{\end{eqnarray}}
\newcommand{\bec}{\begin{cases}}
\newcommand{\eec}{\end{cases}}

\DeclareMathOperator{\Tr}{Tr}

\DeclareMathOperator{\Sgn}{Sgn}
\DeclareMathOperator\arctanh{arctanh}
\DeclareMathOperator\arcoth{arcoth}

\DeclareMathOperator\arcosh{arcosh}

\newcommand{\g}[1]{g_{(#1)}}

\newcommand{\Aa}{\bar{A}}
\newcommand{\OO}{\bar{\Omega}}
\newcommand{\PP}{\bar{P}}
\newcommand{\QQ}{\bar{Q}}

\title{On Acceleration in Three Dimensions}

\author[a]{Gabriel Arenas-Henriquez}
\author[b,c]{Ruth Gregory}
\author[a]{Andrew Scoins}

\affiliation[a]{Centre for Particle Theory, Department of Mathematical Sciences, 
Durham University, South Road, Durham, DH1 3LE, UK}
\affiliation[b]{Theoretical Particle Physics and Cosmology Group, Department of Physics,
King’s College London, University of London,
Strand, London, WC2R 2LS, UK}
\affiliation[c]{Perimeter Institute, 31 Caroline Street North, Waterloo, 
ON, N2L 2Y5, Canada}

\emailAdd{gabriel.arenas-henriquez@durham.ac.uk}
\emailAdd{ruth.gregory@kcl.ac.uk}

\abstract{
We go ``back to basics'', studying accelerating systems in $2+1$ AdS 
gravity \textit{ab initio}. We find three classes of geometry, which we interpret 
by studying holographically their physical parameters. 
From these, we construct stationary, accelerating point particles;
one-parameter extensions of the BTZ family resembling an accelerating black hole;
and find new solutions including a novel accelerating ``BTZ geometry'' not 
continuously connected to the BTZ black hole as well as some black funnel solutions.
}

\begin{document}
\maketitle

\section{Introduction}
\label{sec:intro}

The four-dimensional C-metric \cite{Kinnersley:1970zw}
has been historically considered as the prototypical model of an accelerating black hole.
It consists of a black hole horizon, undergoing uniform acceleration,
isolated from a second black hole by a non-compact acceleration horizon.
The force of acceleration is provided by a tension (positive or negative)
along the axis of symmetry of the black hole, interpreted as a
codimension-two topological defect (or defects), connecting the black hole 
horizon either to a second black hole or to infinity.

Similar solutions of the Einstein equations exist
in the presence of a cosmological constant \cite{Podolsky:2002nk,Dias:2002mi}
and with rotational and electromagnetic charges
\cite{Plebanski:1976gy,Griffiths:2005qp}.
For $\Lambda<0$, such asymptotically (locally) anti-de Sitter solutions exhibit
a ``slowly accelerating'' phase \cite{Podolsky:2002nk} in which there is neither
an acceleration horizon nor a second black hole, the black hole is simply suspended
at a fixed distance from the centre of AdS -- a locally accelerating frame.
The drastic simplification which occurs in this phase has led to significant 
advances understanding the role of acceleration.
In particular, the consistency of thermodynamic laws
for slowly accelerating black holes has recently been established
\cite{Appels:2017xoe,Appels:2016uha,Anabalon:2018ydc,Anabalon:2018qfv}.
First laws of thermodynamics have also been established for ``rapidly accelerating'' 
black holes \cite{Gregory:2020mmi}, which possess an acceleration horizon, see also
\cite{Ball:2020vzo}.
In these constructions, the thermodynamic tension of the defect
is promoted to a thermodynamic charge \cite{Gregory:2019dtq},
with the conjugate quantity being the defect worldvolume
\cite{Krtous:2019fpo} renormalised in a suitable manner.

Despite these advances in understanding the classical aspects of acceleration,
relatively little is known about the quantum properties 
of the accelerating black hole.
There are two apparent approaches towards addressing this shortcoming.

The first is to exploit the holographic correspondence,
attempting to understand the quantum gravitational system
via its dual description as a quantum field theory \cite{Maldacena:1997re}.
It has recently been shown that particular supersymmetric C-metrics
may be realised as Kaluza-Klein truncations of eleven-dimensional supergravity
solutions devoid of topological defects \cite{Ferrero:2020twa}.
These uplifted solutions give a first step towards
a direct holographic understanding of the C-metric.
See also \cite{Dehghani:2001ft,Dias:2013bwa}, for studies investigating
the holographic data available from conical defects which have been ``smoothed''
by matter fields.

The second approach is to exploit the simplification of general relativity in
dimensions fewer than four, for which the theory is topological \cite{Deser:1983nh}.
Lower-dimensional gravity still exhibits many of the interesting features
shared by its higher-dimensional sibling, including the admission of black holes
\cite{Banados:1992wn,Banados:1992gq}.
However, from the perspective of the path integral,
the lack of dynamics renders the theory radically simpler \cite{Witten:1988hc}.
In fact, for asymptotically AdS solutions (without defect singularities)
the path integral may be evaluated explicitly \cite{Maloney:2007ud}.
By constructing three-dimensional solutions analogous to the C-metric,
one may be able to apply similar technology
to gain insight into the nature of acceleration.
It is this latter approach we follow in this paper.

Though we have motivated the C-metric as a description of an accelerating black hole,
more generally one may select less typical ranges of parameters in the C-metric 
to find a menagerie of solutions \cite{Hubeny:2009kz}.
In this way, one can construct geometries with horizons that extend to the conformal boundary;
they are dual to strongly coupled field theories living on black hole backgrounds.
Such solutions are known as either black funnels or black droplets \cite{Hubeny:2009ru}.
In fact, the acceleration horizon formed in the rapidly accelerating phase of the AdS C-metric
is one such droplet.
In this paper we demonstrate that an analogous collection of solutions exists in 
three dimensions, providing a range of black funnel and droplet solutions.
This builds on the work of \cite{Astorino:2011mw}, which presented both a solution 
describing an accelerating conical deficit, or point particle, \cite{Anber:2008zz},
and a solution analogous to the C-metric describing a BTZ black hole \cite{Banados:1992wn}
with a codimension-one defect emerging from its horizon.
This defect is under compression so is commonly referred to as a ``strut'' as
it possesses a negative energy density.
We show that not only can a more physical solution be constructed,
describing a BTZ black hole accelerated by a positive tension wall, but 
the solution with a strut exhibits a richer phase structure than previously
noticed in the literature,
possessing a ``rapid phase'' in which a disconnected black droplet forms.
This phase transition to a droplet solution is directly analogous
to the formation of an acceleration horizon in the four-dimensional theory.
We also construct a solution analogous to the C-metric describing a white hole
and one exhibiting a black funnel.
We calculate the masses of our solutions using holographic techniques,
and comment on their other physical properties.

The paper is organised as follows.
In section \ref{sec:solutions} we outline the process for constructing solutions
from a C-metric-like ansatz, categorising the resultant possibilities into three classes.
In sections \ref{sec:accptcle} we construct and analyse accelerating particles, 
which are represented as conical defects in the AdS geometry.
In section \ref{sec:IIstrut} we clarify the results of \cite{Astorino:2011mw},
discussing their BTZ-like solution with a domain wall under compression.
We show that the solution exhibits a more broad range of behaviours
than previously acknowledged,
and call into question the author's conclusion that this is
an ``accelerating'' BTZ black hole. We then discuss a similar, yet more physical,
family of black holes possessing instead a domain wall under tension.
In section \ref{sec:IC} we construct a novel white hole solution,
with a domain wall defect under tension, analogous to the C-metric.
We construct a novel three-dimensional black funnel in section \ref{sec:III},
and conclude in section \ref{sec:discussion}.

\section{C-metric-like solutions in three dimensions}
\label{sec:solutions}

We start with an ansatz similar to the typical form of the four-dimensional C-metric
\cite{Plebanski:1976gy,Podolsky:2002nk}, as in \cite{Astorino:2011mw,Xu:2011vp}, 
\begin{equation}
    ds^2 = \frac{1}{\OO^2}
      \Big[
        \PP(y)d\tau^2
        -\frac{dy^2}{\PP(y)}
        -\frac{dx^2}{\QQ(x)}
      \Big] \,,
\end{equation}
with the conformal factor given by
\begin{equation}
  \OO = \Aa(x-y) \,.
  \label{conformalfactor}
\end{equation}
Here, $\Aa$ is a parameter with dimensions of inverse-length (the coordinates are
dimensionless). The conformal boundary lies at $x=y$.
The Einstein equations with negative cosmological constant $\Lambda=-\ell^{-2}$
are explicitly solvable, with solution
\begin{equation}
  \QQ(x) = c + bx + ax^2  
  \,,
  \qquad
  \PP(y) = \frac{1}{\Aa^2\ell^2}-\QQ(y)
  \,. 
  \label{metricfunc}
\end{equation}
All three parameters $a$, $b$ and $c$ are, in principle,
arbitrary real numbers, provided the correct signature is retained.
However, there exist gauge redundancies which allow us to eliminate two of the three.
Firstly, eliminate $b$ by noting that the metric is invariant
under translation:
\begin{equation} 
  x     \rightarrow x+s     \,,\qquad
  y     \rightarrow y+s     \,.
\end{equation}
Then, noting the existence of a dilatation symmetry:
\begin{equation}
  \tau \rightarrow s \tau  \,,\qquad
  x    \rightarrow s x     \,,\qquad
  y    \rightarrow s y     \,,
\end{equation}
the effect of which is to scale $a$ by a positive factor $s^2$, means we 
have the freedom to choose $|a|=|c|$.
There are then two possibilities --
either the signs of $a$ and $c$ are the same, or they differ.
The upshot of this discussion is that the parameter space
consists of $3$ possible classes of geometry\footnote{The
fourth option, with both $c<0$ and $\Delta_{\QQ}<0$,
has no range of $x$ giving a Lorentzian signature, so
we disregard this case.}
distinguished by the sign of $c$ and that of the discriminant 
$\Delta_{\QQ} =b^2-4ac$.
Their definitions are given in table \ref{tab:1}. 
\begin{table}[b]
\centering
\begin{tabular}{ | c || c | c | c |}
\hline 
Class   & $\Sgn \Delta_{\QQ}$ & $\Sgn c$ \\
\hline\hline 
I       & $+$       & $+$     \\
II      & $+$       & $-$     \\
III     & $-$       & $+$     \\  
\hline
\end{tabular}
\caption{The three classes of solution and their defining characteristics.}
\label{tab:1}
\end{table}
There is also a symmetry under parity
\begin{equation}
x     \rightarrow -x    \,,\qquad
y     \rightarrow -y    \,,
\end{equation}
that allows us to take $x>y$.
This is the same as a conventional choice one makes for
the four-dimensional C-metric\footnote{However,
in four dimensions there is an equivalent alternative:
the symmetry for the four-dimensional metric is
$x\rightarrow -x, y\rightarrow -y,\mu\rightarrow-\mu$, 
where $\mu$ is the coefficient of the $y^3$ term now present in $P(y)$.
This has led some authors \cite{Hubeny:2009kz}
to fix $\mu>0$, which leaves a choice of orderings for $x$ and $y$.
We do not have such an option here since $P$ is quadratic.
}.

Finally, a simple parameter redefinition $A=\sqrt{|c|}\Aa$ 
shifts all of the dependence on $c$ into the timelike component of the metric
which, since the geometry is static, 
we are free to absorb by a rescaling $\tau\rightarrow \tau/|c|$.
We have arrived at a set of \textit{canonical gauges}
in which all of the dependence on $a$, $b$, and $c$ has been removed.
The metric in this gauge takes the form
\begin{equation}
ds^2 = \frac{1}{\Omega^2} \Big[ P(y)d\tau^2 
-\frac{dy^2}{P(y)} -\frac{dx^2}{Q(x)} \Big] \,,
\label{eq:metricxy2}
\end{equation}
with the conformal factor given by
\begin{equation}
\Omega = A(x-y)
\label{conformalfactor2}
\end{equation}
and the metric functions in table \ref{tab:3}.
Also given in table \ref{tab:3} are the maximal ranges of $x$ for which 
$Q(x)$ is positive, thus yielding a Lorentzian metric.
\begin{table}[t!]
\centering
\begin{tabular}{ | c || c | c | c | }
\hline
Class   & $Q(x)$    & $P(y)$    & Maximal range of $x$ \\
\hline\hline 
I       & $1-x^2$   & $\frac{1}{A^2\ell^2}+(y^2-1)$ & $|x|<1$ \\
II      & $x^2-1$   & $\frac{1}{A^2\ell^2}+(1-y^2)$ & $x>1$ or $x<-1$ \\
III     & $1+x^2$   & $\frac{1}{A^2\ell^2}-(1+y^2)$ & $\mathbb{R}$ \\
\hline 
\end{tabular}
\caption{The metric functions in canonical gauge,
together with the largest available range of $x$.}
\label{tab:3}
\end{table}

It is worth pausing a moment to consider the solutions thus obtained. 
Class I solutions have a familiar form, they look like the 
4D Rindler metric written in ``C-metric'' coordinates; identifying $x = \cos\theta$
gives us a polar system of coordinates that has an origin if $y\to-\infty$.
Classes II and III look somewhat different, as $y$ is now bounded, and $x$ no
longer looks analogous to an angular coordinate. We will see that Class I 
has the same analogies to slow and rapid acceleration as in the 4D case, but 
also admits a novel solution in a corner of parameter space with the interpretation
of accelerating BTZ black hole. This solution is parametrically disconnected
from the BTZ black hole (i.e.\ cannot be obtained by letting $A\to0$ continuously)
and is a new solution not previously noted in the literature.
Class II represents accelerating BTZ solutions continuously connected to 
the BTZ black hole. Class III is more analogous to a braneworld. 

We can also sub-divide the classes further by their causal or horizon structure, 
directly related to the value of $A\ell$. Just as with their 4D cousins, class I solutions 
fall into three possible categories: those which satisfy $A^2\ell^2<1$ lack event 
horizons, those with $A^2\ell^2>1$ possess two distinct event horizons, and 
solutions with $A\ell=1$ possess a single event horizon at $y=0$. We christen these 
subclasses \textit{slow}, \textit{rapid}, and \textit{saturated} respectively. 
Solutions of Class II have a persistent horizon structure over all values of $A$ 
and $\ell$. However, they may be instead categorised by the choice to take $x$ 
to live in a particular connected component, with either $x<-1$ or $x>1$. We 
christen these the \textit{left} and \textit{right} subclasses of II respectively. 
Metrics of Class III attain $P$ positive only if $A^2\ell^2<1$. There are then 
two event horizons, with the valid range of $y$ between them.

So far, none of these solutions appear to have a mass parameter, this is
because there is no propagating mode of the graviton in three dimensions.
Any ``mass'' is generated by some sort of identification of pure 3D AdS: 
either one creates a conical deficit in the spacetime, with the interpretation of 
the deficit as a point mass, or one quotients the spacetime, with the
interpretation of the BTZ black hole. Apart from identifications, the local 
spacetime is precisely 3D AdS, thus all these three classes are simply different 
patches of AdS coordinatised in a variety of ways. We derive the coordinate
transformations in appendix \ref{app:map}, but for now, note that all of these metrics with
$x$ taking its maximal range are just patches of AdS, and do not per se 
have an interpretation as an accelerating black hole.

In order to have the interpretation of a point mass or black hole, we must perform
the relevant identifications on the geometry. In each case, this involves cutting the 
spacetime along an $x=$const.\ curve, and identifying with another equivalent
$x=$const.\ curve. Because of the off-centre coordinates, this will turn out to
introduce a domain wall into the geometry -- a codimension 1 defect that
provides the necessary force for acceleration.
 
Having identified the largest available range of $x$ in each solution class,
we note that (for classes I and II)
as $|x| \to 1$, the metric is regular and the surface $|x|=1$ has 
vanishing extrinsic curvature. It is therefore natural to think of $x$ as an
angular variable and to identify two mirror images of our geometry across $|x|=1$.
If we then cut each geometry across another constant $x-$surface, $x = x_0$,
this will introduce a domain wall into the system, with a tension, $\sigma$,
given by the Israel equations \cite{Israel:1966rt}
\be
4\pi \sigma \, h_{ij} = 4\pi \int_-^+ T_{ij} = K_{ij} - K h_{ij}   \,
\ee
where $K_{ij}$ is the extrinsic curvature of the $x=x_0$ surface.
The (inward pointing) unit normal to $x=x_0$ is
\begin{equation}
\textbf{n} = \pm\Omega^{-1}Q^{-\frac{1}{2}}dx \,,
\label{eq:norm}
\end{equation}
where the plus sign is taken if $x>x_0$ in the interior (i.e.\ if $x \in [x_0,1]$ for class I,
$x > x_0$ for class II), and vice versa. The same principle applies for class III 
solutions, however, there is now 
no surface of zero extrinsic curvature, thus we have the choice of having a single
cut at $x=x_0$, or a second cut at some other value of $x$. This is precisely
analogous to the Randall-Sundrum models 2 and 1 
\cite{Randall:1999vf,Randall:1999ee} respectively.
The extrinsic curvature $K_{ij}=\nabla_in_j$ is then easily computed as
\begin{equation}
K_{ij} = \mp A \sqrt{Q}h_{ij} \,,
\end{equation}
leading to a domain wall defect with tension
\begin{equation}
\sigma=\pm\frac{A}{4 \pi }\sqrt{Q(x)} \;.
\label{eq:tension}
\end{equation}
This now demonstrates explicitly that taking a mirror image copy of
our spacetime and gluing across $|x|=1$ leads to a continuous and smooth
geometry, whereas gluing across $x=x_0$ leads to a defect with 
positive tension if $x<x_0$, or negative tension if $x>x_0$ lies within
our spacetime. For convenience, and to easily distinguish between positive
and negative tension defects, we call the positive tension defect a {\bf (domain) wall},
and the negative tension one a {\bf strut}.
\begin{figure}[tb!]
\centering
\includegraphics[width=0.49\textwidth]{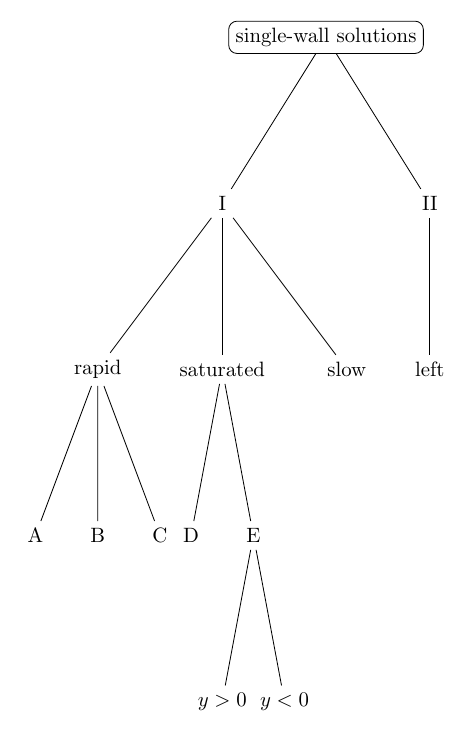}~
\includegraphics[width=0.5\textwidth]{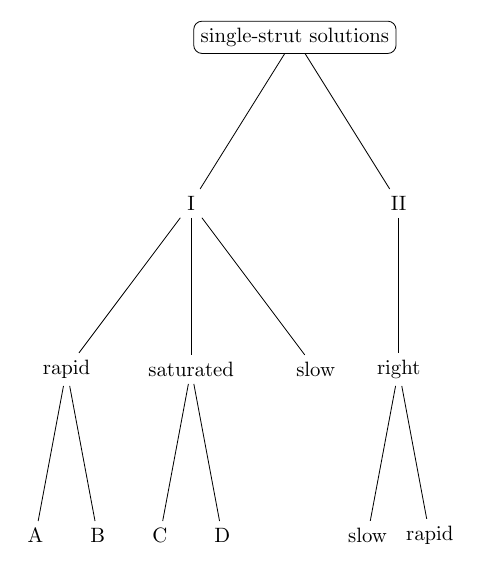}
\caption{Classification of distinct single-wall and single-strut solutions with $t$ timelike.
Each leaf node represents a qualitatively-distinct solution.}
\label{fig:classificationtree}
\end{figure}

When restricting the range of the $x$ coordinate, 
it may be that the new range of parameters ``avoids''
one or more of the event horizons that would have been present,
or changes the number of horizons that meet the conformal boundary,
had the full range of $x$ been considered.
As a consequence, (with $t$ timelike) there are
eight qualitatively distinct single-wall and
seven qualitatively distinct single-strut solutions
one can construct through various choices of $x_0$
shown in figure \ref{fig:classificationtree}.
\begin{figure}
\centering
\includegraphics[width=0.33\linewidth]{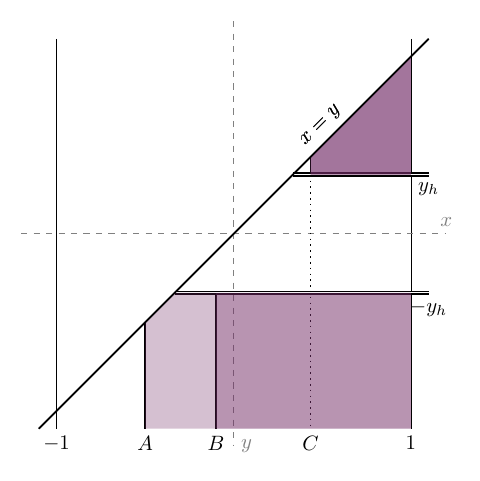}~
\includegraphics[width=0.33\linewidth]{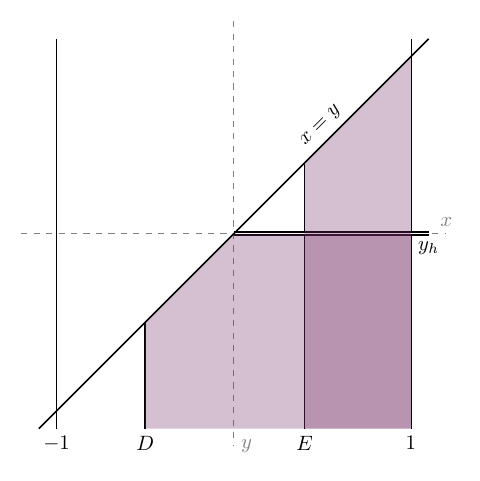}~
\includegraphics[width=0.33\linewidth]{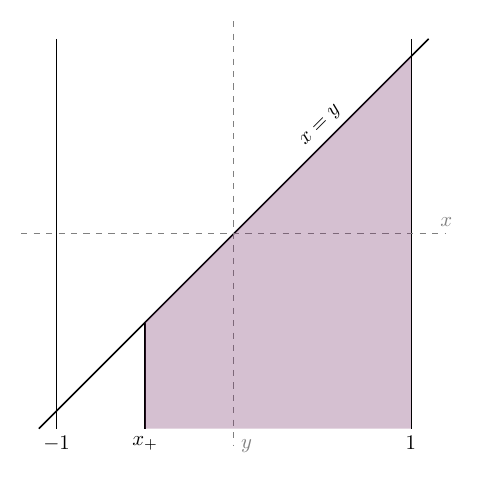}
\caption{Coordinate ranges for single-wall solutions constructed from 
metrics of Class I (with $t$ timelike). Left: three qualitatively distinct 
solutions are shown for class I\textsubscript{rapid}, 
Centre: three for class I\textsubscript{saturated}, and Right: one 
for class I\textsubscript{slow}.
}
\label{fig:I}
\end{figure}%

\begin{figure}
\includegraphics[width=0.35\linewidth]{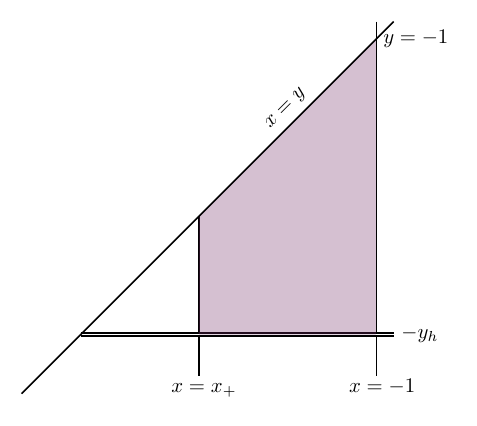}
\hskip 2cm
\includegraphics[width=0.33\linewidth]{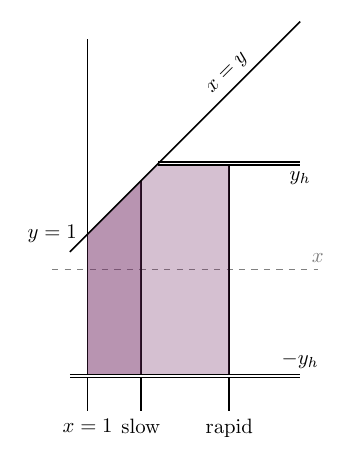}
\caption{Coordinate ranges for single-defect solutions constructed from 
metrics of Class II. Left: A wall solution for class II\textsubscript{left}, and
Right: two qualitatively distinct strut solutions for class II\textsubscript{right}.
  }
\label{fig:II}
\end{figure}

In slightly more detail, figure \ref{fig:I} shows the coordinate patching for
all of the possible distinct single-wall solutions in Class I geometries.
Figure \ref{fig:I}(Left) shows
the rapidly accelerating subclass I\textsubscript{rapid},
with three possible single-wall solutions:
$A$ denotes a choice of $x_+\in(-1,-y_h)$ with $y<0$,
for which there are no horizons.
$B$ denotes a choice of $x_+\in(-y_h,1)$ with $y<0$,
for which there is a single horizon.
$C$ denotes a choice of $x_+\in(-y_h,1)$ with $y>0$.
There are three similar solutions for I\textsubscript{saturated},
shown in figure \ref{fig:I}(Centre):
$D$ denotes a solution with $x_+<0$ and $E$ one with $x_+>0$.
Having chosen $x_+>0$, one may then construct single-wall solutions 
with either $y>0$ or $y<0$. Figure \ref{fig:I}(Right) corresponds to
the slowly accelerating configuration with a domain wall.
To complete the solutions with a domain wall, figure \ref{fig:II}(Left) 
shows the unique single-wall solution derived from Class II.

\begin{figure}
\centering
\includegraphics[width=0.33\linewidth]{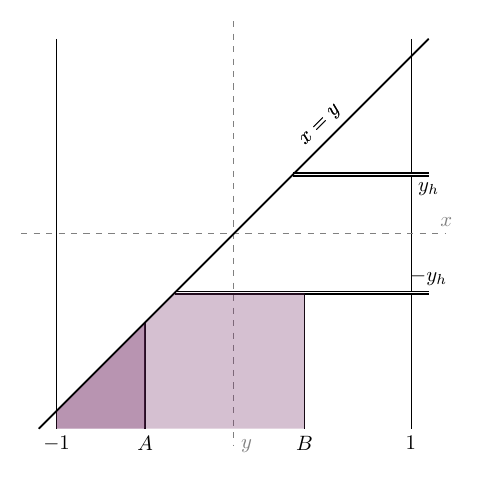}~
\includegraphics[width=0.33\linewidth]{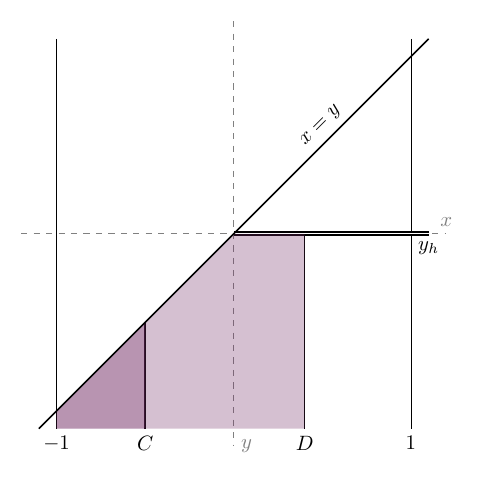}~
\includegraphics[width=0.33\linewidth]{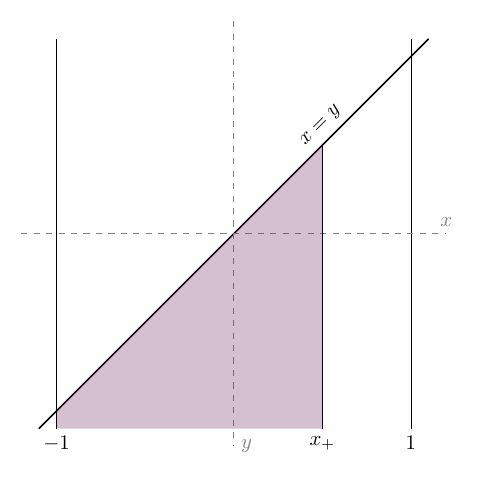}
\caption{Coordinate ranges for single-strut solutions constructed from metrics of Class I
(with $t$ timelike). Left: two qualitatively distinct solutions are shown for class
I\textsubscript{rapid}, Centre: two for class I\textsubscript{saturated}, and Right:
one for class I\textsubscript{slow}.
}
\label{fig:Istrut}
\end{figure}

Similarly, figure \ref{fig:II}(Right) shows the unique single strut Class II solution,
and figure \ref{fig:Istrut}
shows all of the possible qualitatively distinct single-strut solutions
constructable from Class I geometries.
The letters $A$, $B$, $C$, and $D$
in figures \ref{fig:Istrut}(Left) and \ref{fig:Istrut}(Centre)
denote choices of $x_+$ that determine whether the constructed solution
has (for $B$ and $D$), or does not have (for $A$ and $C$), a horizon.
Figure \ref{fig:Istrut}(Right) corresponds to
the slowly accelerating configuration with a strut.

We now turn to the physics of each
solution class, describing the various spacetimes and discussing their
holographic interpretations.

\section{An Accelerating Particle}
\label{sec:accptcle}

A point particle in 3-dimensional gravity is represented by a conical deficit
-- a $\delta$-function in the Ricci curvature. This means we have to have an 
``origin'' in order to cut out the deficit. A quick look at the metric
\eqref{eq:metricxy2} shows that the length of an angular arc, represented by $x$,
becomes zero if $y\to-\infty$ (recall $y<x$). Table \ref{tab:3} then
shows that only Class I solutions are able to satisfy this requirement
while keeping $P\geq0$.

To construct the accelerating conical defect, we first glue two
copies of the Class I geometry \eqref{eq:metricxy2} along $x = \pm1$,
so that we can identify an angular coordinate via $x = \pm \cos\theta$.
We then introduce a conical deficit by choosing an $x_+\in(-1,1)$ 
and restricting the range of $x$ to $\pm x\in(x_+,1]$.
The $x=\pm1$ axis of the newly formed spacetime is regular,
but the origin at $y\to-\infty$ has an angular deficit of $2[\pi - \arccos(\pm x_+)]$,
with the $x=x_+$ axis marking the position of a wall of
tension $\sigma=\pm A\sqrt{Q(x_+)}/{4\pi}$.
\begin{figure}[t!]
\centering
\includegraphics[width=0.5\textwidth]{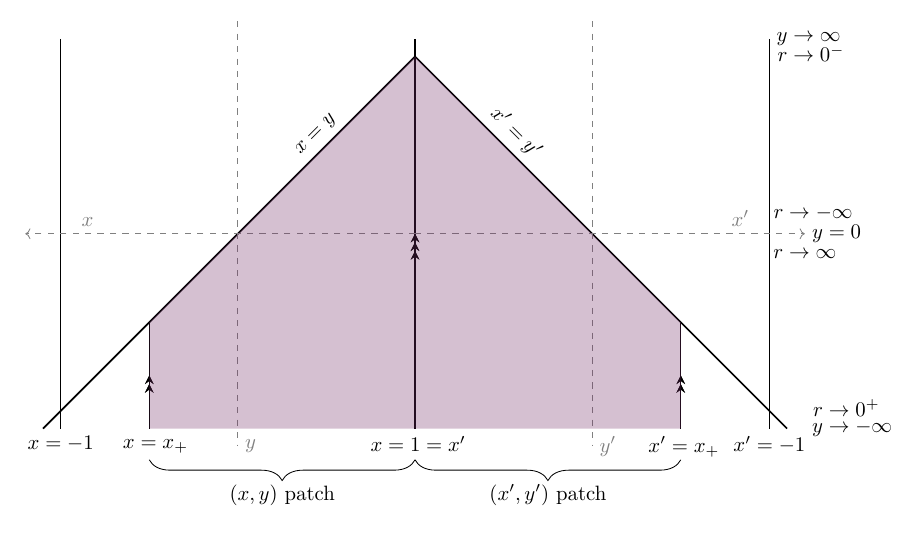}~
\includegraphics[width=0.5\textwidth]{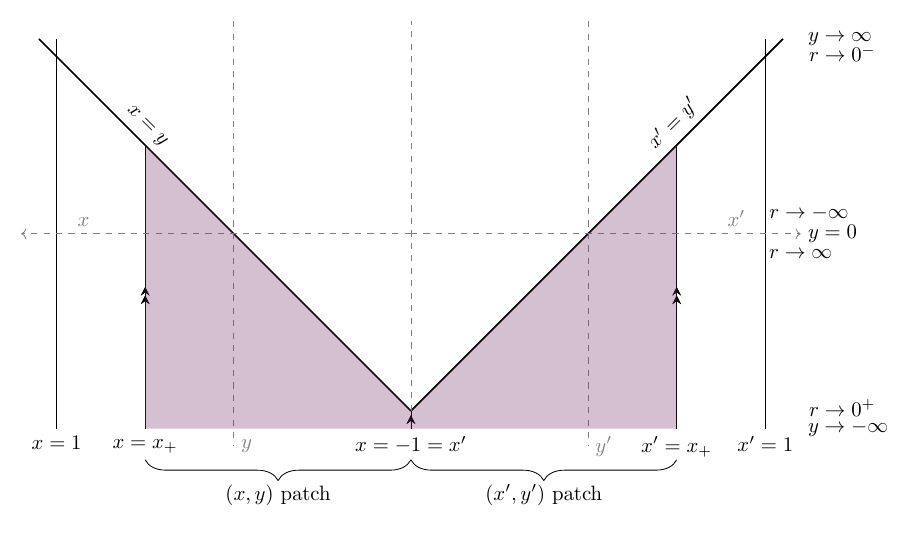}
\caption{The two patches of type I spacetime
used to construct a slowly accelerating conical defect.
{\bf Left:} Point particle pulled by a domain wall.
{\bf Right:} Point particle pushed by a strut-wall. }
\label{fig:IslowGlued}
\end{figure}

We now focus on the positive tension case ($x\in(x_+,1]$), describing
in detail the accelerating particle pulled by a domain wall, briefly 
concluding this section with the negative tension strut, which 
follows largely the same calculational procedure.

\subsection{A particle pulled by a domain wall}
\label{sec:Istring}

For class I geometries, it is useful to rewrite the coordinates in a more
natural form that capitalises on our $(3+1)$-dimensional intuition. 
We introduce dimensionful coordinates $r=-(Ay)^{-1}$ and $t=\alpha A^{-1}\tau$,
and identify $x$ with an angular coordinate via $x=\cos(\phi/K)$.
The parameter $K\equiv\pi/\arccos(x_+)$ now encodes the range of the 
$x$ coordinate, fixing the range of $\phi$ to be $(-\pi,\pi)$.
The metric is now
\be
\beal
ds^2 &= \frac{1}{\left[1+Ar\cos\left(\phi/K\right)\right]^2}
\left[ f(r)\frac{dt^2}{\alpha^2}  -\frac{dr^2}{f(r)}
-r^2 \frac{d\phi^2}{K^2} \right] \,, \\
f(r) &= 1 + (1-A^2\ell^2)\frac{r^2}{\ell^2} \,.
\eeal
\label{eq:classIrphi}
\ee
Note that with foresight, we have left explicit the possibility of rescaling $t$ 
by some dimensionless parameter $\alpha$!

In this form, we now have a natural $A\to0$ limit recovering pure AdS,
as well as being able to identify slow ($A\ell<1$), saturated ($A\ell=1$),
and rapid ($A\ell>1$) acceleration readily. For $A\ell<1$, no horizons are present, 
and we will see that the accelerating particle is, analogous to the 4D case, 
a conical deficit suspended are fixed distance from the centre of AdS.
As $A$ gets larger ($A\geq1$), a Killing horizon forms at $r = \ell/\sqrt{A^2\ell^2-1}$
which we refer to as an \textit{acceleration horizon}. Provided the acceleration is
not too large (see below) this horizon is non-compact and masks
part of the conformal boundary. The domain wall extends between the particle 
and the boundary. However, for large acceleration, the domain wall extends from 
the particle to the horizon, and after identification, the horizon `wraps around' to 
mask the entire conformal boundary and the spatial sections become compact.

The polar-style coordinate system is helpful in interpreting the accelerating 
particle spacetimes as the conical deficit sits at $r=0$ and is easily visualised.
Irrespective of acceleration, this is achieved by having $K>1$ in the metric
\eqref{eq:classIrphi}. The angular deficit, which one may be inclined to refer 
to as the ``particle mass''
\cite{Deser:1983tn}, is
\begin{equation}
  m_c=\frac{1}{4}\left(1-\frac{1}{K}\right)
  \,,
  \label{eq:Istringdeficit}
\end{equation}
while the tension in the domain wall is
\begin{equation}
  \sigma = \frac{A}{4\pi}\sin{\left(\frac{\pi}{K}\right)}
  \,.
  \label{eq:Istringtension}
\end{equation}
When $K=1$, both the conical deficit \eqref{eq:Istringdeficit}
and domain wall tension \eqref{eq:Istringtension} vanish,
and the geometry becomes global AdS$_3$ written 
in an accelerated frame.
We have monotonically increasing particle mass as $K$ increases,
whereas the domain wall tension first increases, reaching a maximum 
of $A/4\pi$ at $K=2$, then decreases further with increasing $K$.
As $K\rightarrow\infty$, the ``particle mass'' asymptotes $1/4$,
and the domain wall tension tends to zero. In this case, we have 
excised all of the spacetime, and one can view this as there being
nothing to accelerate!
As $A\rightarrow 0$, the `mass' of the particle is unaffected
and the tension of the domain wall goes to zero.

In the rapid or saturated phases, it is the magnitude of the particle 
mass relative to the acceleration that 
determines the horizon structure.
For $4\pi \sigma \ell<1$ (or $x_+<-y_h$), 
the acceleration horizon is non-compact and the domain wall 
reaches the boundary.
For strongly accelerating particles $A\ell \sin(\pi/K)>1$,the acceleration horizon compactifies
and we have a de Sitter-like state.

To identify that the conical defect is accelerating,
consider the four-acceleration along the worldline traced by the origin.
The origin $r=0$ of the coordinates \eqref{eq:classIrphi}
has normalised four-velocity
\begin{equation}
  \mathbf{u} =
  \frac{\alpha\Omega}{\sqrt{f}}
  \Bigg\vert_{r\rightarrow 0}
  \partial_t
  \,.
\end{equation}
The associated four-acceleration is then given by a covariant derivative,
\begin{equation}
  \mathbf{a} = \nabla_\mathbf{u} \mathbf{u}
  \,,
\end{equation}
which has magnitude
\begin{equation}
  \left\vert\mathbf{a}\right\vert
  =
  \sqrt{-\left(\nabla_\mathbf{u} \mathbf{u}\right)^2}
  =
  A
  \,.
\end{equation}
We see that the acceleration parameter gives the locally experienced 
acceleration of the particle.

One can also identify the particle's acceleration by considering
the temperature of the horizon in the rapid phase.
In this phase, demanding regularity of the Euclidean section
indicates a horizon temperature of 
\begin{equation}
  T = \frac{A}{2\pi\alpha}\sqrt{1-\frac{1}{A^2\ell^2}}
  \,.
  \label{eq:classItemperature}
\end{equation}
For accelerations which are large
compared to the AdS scale,
$A\gg \ell^{-1}$,
the particle is close to the acceleration horizon and provides 
effects which dominate the effect of the background curvature.
In this regime, the geometry is approximately flat,
and one should expect $\alpha\sim 1$.
The horizon is then seen to be Rindler, with temperature $T=(2\pi)^{-1}A$.

We now wish to identify the correct mass associated with this solution.
This is typically tricky in the presence of acceleration horizons,
though we refer the reader to \cite{Gregory:2020mmi}
for some recent progress.
To sidestep the issue, focus on the slowly accelerating phase,
$A^2\ell^2<1$, which possesses no horizons.
In order to identify the correct mass,
one must normalise the generator of time translations $\partial_t$
to that of an observer located at the boundary \cite{Caldarelli:1999xj}.
That is, we wish to choose $\alpha$ in \eqref{eq:classIrphi}
such that $t$ is the same time coordinate as when the geometry is written 
in global gauge
\begin{equation}
  ds^2=
  \left(1+\frac{R^2}{\ell^2}\right) dt_{\text{Global}}^2
  - \frac{dR^2}{\left(1+\frac{R^2}{\ell^2}\right)}
  - R^2 d\vartheta^2
  \,.
  \label{eq:GlobalAdS3}
\end{equation}
Note that the function $f$ has the same form as the comparable function 
in the C-metric in four dimensions,
in the limit of vanishing mass parameter \cite{Griffiths:2006tk}.
Hence, the transformation between local and global coordinates
is already known to be \cite{Podolsky:2002nk}
\begin{equation}
  \left(1+\frac{R^2}{\ell^2}\right) =
  \frac{f(r)}{\alpha^2\Omega(r,\phi)^2}
  \,,
  \qquad
  R\sin\vartheta =
  \frac{r\sin{\left(\phi/K\right)}}{\Omega(r,\phi)}
  \,.
\label{eq:ILocaltoGlobal}
\end{equation}
This holds only if the temporal rescaling is 
\begin{equation}
  \alpha=\sqrt{1-A^2\ell^2}
  \,.
  \label{eq:Ialpha}
\end{equation}

\begin{figure}[t!]
  \centering
  \includegraphics[width=0.5\textwidth]{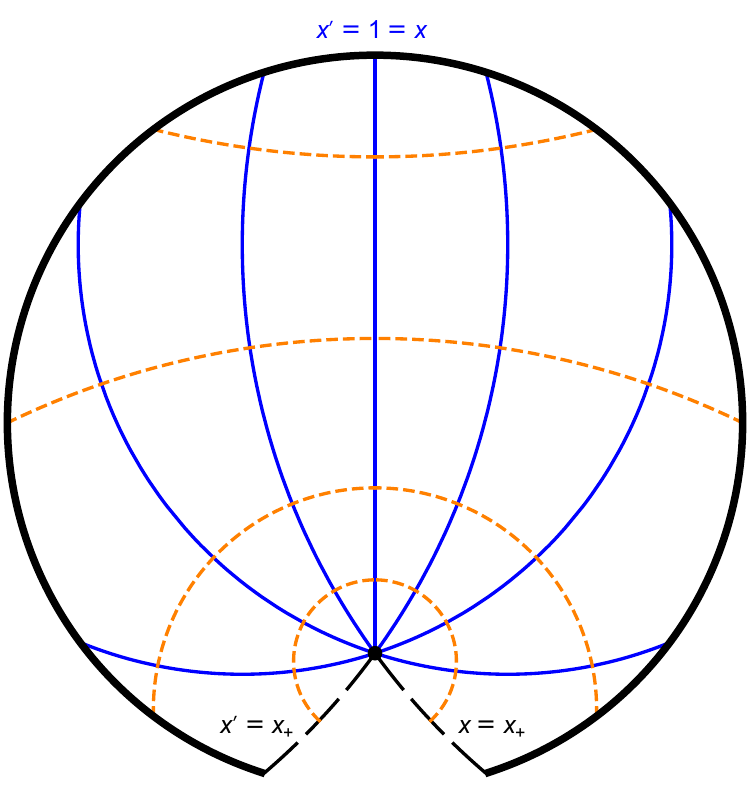}
  \caption{
      The slowly accelerating conical deficit with $A=0.9\ell$,
      pulled by a wall, mapped onto the Poincar\'{e} disk.
      The deficit is shown as a black point and accelerates southwards.
      A ``wedge'' is removed from the global space, with its edges
      $x=x_+$ and $x'=x_+$
      (long-dashed black) identified to attain a wall.
      Several lines of constant $x$ are shown in blue,
      with lines of constant $y$ in dashed orange.}
  \label{fig:3DIstringslow}
\end{figure}

This mapping to global gauge allows for an intuitive picture of the spacetime 
to be constructed.
By compactifying onto the Poincar\'{e} disk,
we obtain figure \ref{fig:3DIstringslow}
for the slowly accelerating conical deficit,
which is plotted at some fixed time $t$.
The details of the compactification are given in appendix \ref{app:map}.
One must use two $(x,y)$ patches
to cover both the eastern and western hemicircles,
as in figure \ref{fig:IslowGlued}.
Note that, since $t$ aligns with the global time coordinate,
one may stack an infinite number of identical diagrams to visualise
the complete space.
We find a conical deficit ``pulled'' closer to the boundary than
the origin of global coordinates by the removal of a distorted wedge.

In fact, although defining a mass in the rapidly accelerating case
is difficult,
we can already determine the appropriate Killing vector one should use
once a reliable definition becomes available.
Consider the portion of AdS$_3$ described by the metric
\begin{equation}
  ds^2 = \left(-1+\frac{R^2}{\ell^2}\right) dt_{\text{Rindler}}^2
        -
        \frac{dR^2}{\left(-1+\frac{R^2}{\ell^2}\right)}
        -
        R^2 d\vartheta^2
        \,,
        \label{eq:RindlerWedge}
\end{equation}
with $R>\ell$ and $\vartheta\in \mathbb{R}$.
This geometry possesses a non-compact bifurcate Killing horizon at $R=\ell$,
generated by $\partial/\partial_{t_{\text{Rindler}}}$,
and we will refer to it as either the \textit{planar BTZ geometry} or \textit{Rindler wedge}.
Close to the conformal boundary at large $R$,
$t_{\text{Rindler}}$ is the usual timelike coordinate of the Poincar\'{e} patch.
This time coordinate provides the zero-mass state for AdS$_3$ \cite{Balasubramanian:1999re}.
As such, the appropriate normalisation of $t$ for the rapidly accelerating particle
is given by choosing $\alpha$ such that the time coordinate of the solution \eqref{eq:classIrphi}
matches that of the Rindler wedge \eqref{eq:RindlerWedge}.
This choice of Killing vector closely mirrors the choice made in
\cite{Gregory:2020mmi}, where,
for four-dimensional accelerating black holes without cosmological constant,
the generator of the Rindler horizon was shown to provide a mass satisfying a first law.
The transformation between the rapid Class I and Rindler wedge is given by 
\begin{equation}
  \left(-1+\frac{R^2}{\ell^2}\right) =
  \frac{f(r)}{\alpha^2\Omega(r,\phi)^2}
  \,,
  \qquad
  R\sinh\vartheta =
  \frac{r\sin{\left(\phi/K\right)}}{\Omega(r,\phi)}
  \,,
\label{eq:ILocaltoRindler}
\end{equation}
where we must take
\begin{equation}
  \alpha=\sqrt{A^2\ell^2-1}
  \,.
  \label{eq:IalphaRapid}
\end{equation}

\begin{figure}[t!]
  \centering
  \includegraphics[width=0.45\textwidth]{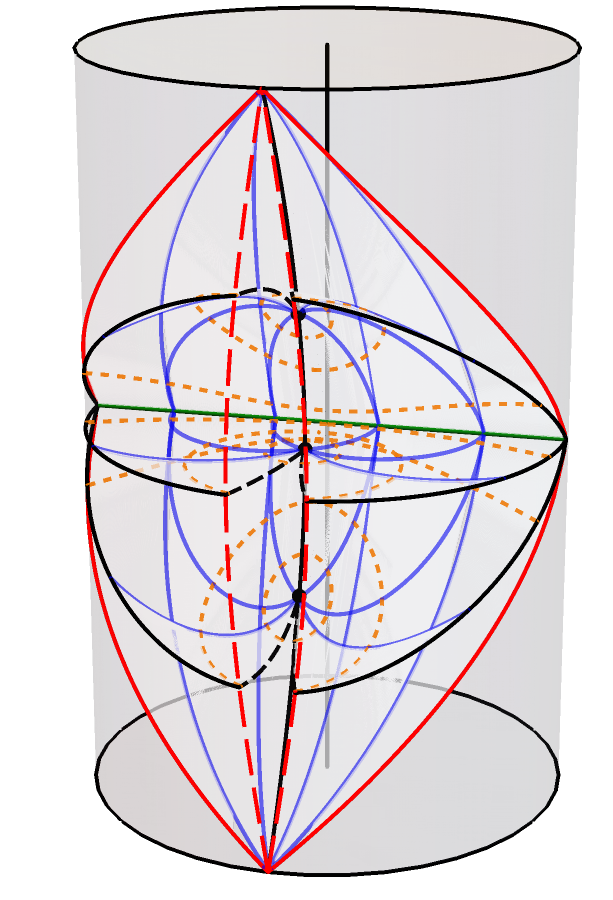}
  \hfill
  \includegraphics[width=0.45\textwidth]{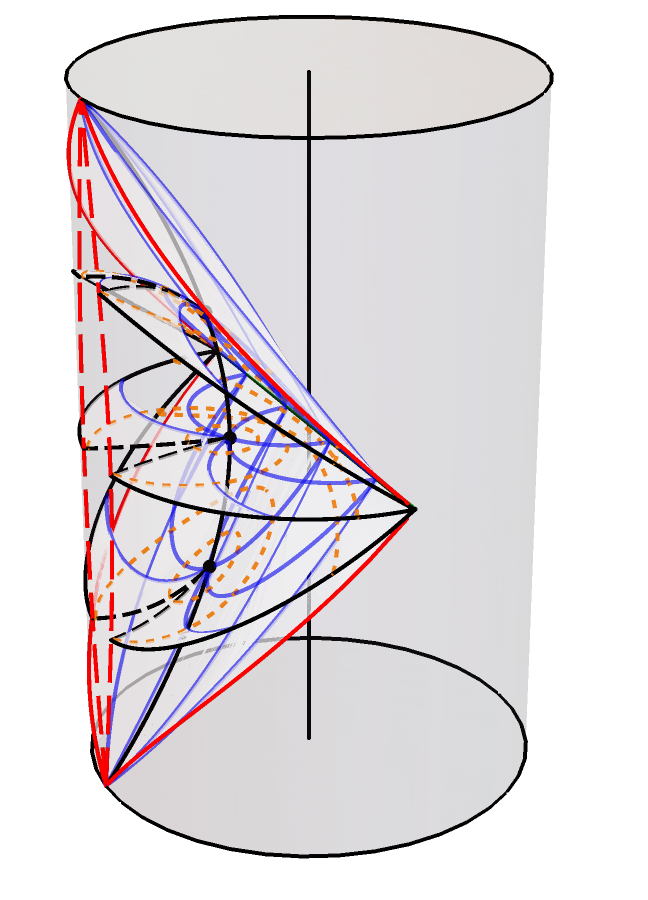}
  \caption{
      The Class I\textsubscript{rapid,A} solution
      embedded within global AdS$_3$.
      The particle worldline is shown in solid black.
      Several surfaces of constant $t$ are plotted.
      The event horizons are demonstrated by the surfaces at early and late $t$.
      The bifurcation surface is shown as a green line.
      A ``wedge'' is removed, with its edges $x=x_+$ and $x'=x_+$ identified;
      this is the location of the domain wall.
      These edges are shown in long-dashed black within each time-slice,
      and in dashed red at the boundary.
      The boundary of the classically accessible
      subset of the global boundary is shown in red.
      Lines of constant $x$ are shown in blue,
      with lines of constant $y$ in dashed orange.
      To guide the eye, the axis of the cylinder is also shown in black.}
  \label{fig:3DIstringrapid}
\end{figure}

Much like the slowly accelerating case,
the transformation between local coordinates and the Rindler wedge
may be used to understand the rapidly accelerating conical deficit
as a subset of AdS$_3$.
For light conical deficits, this space is the planar BTZ or Rindler geometry 
with a wedge removed.
This is shown in figure \ref{fig:3DIstringrapid}. Note a crucial difference with 
the 4D accelerating black hole with an acceleration horizon that has a partner
black hole in the mirror of the accelerating patch: Here, we can cut out a single 
conical deficit, and have only a solo accelerating particle - due to the local AdS
spacetime being exact, there is no mandate for a mirror black hole. This can also
be contrasted with the strut case discussed later (Figure \ref{fig:3DIstrutrapid}).
We must plot two $(x,y)$ patches in order to cover the Rindler wedge.
Details of the mapping onto the global space given in appendix \ref{app:map}.
A key feature is that the Rindler time coordinate and the global time
do not align.
In figure \ref{fig:3DIstringrapid},
global time runs vertically, aligned with the axis of the cylinder.
The conical defect on the other hand is seen to accelerate in
from the conformal boundary at early times $t$, 
then move back out towards the boundary as $t\rightarrow\infty$.
Both of these events happen in finite global time,
resulting in the particle experiencing an acceleration horizon.
It is already known that the planar BTZ geometry describes an acceleration 
horizon for test particles \cite{Banados:1992gq};
our solution describes a particular physical object undergoing acceleration,
with the force of acceleration provided by a physical domain wall.

\begin{figure}[t!]
  \centering
  \includegraphics[width=0.45\textwidth]{%
    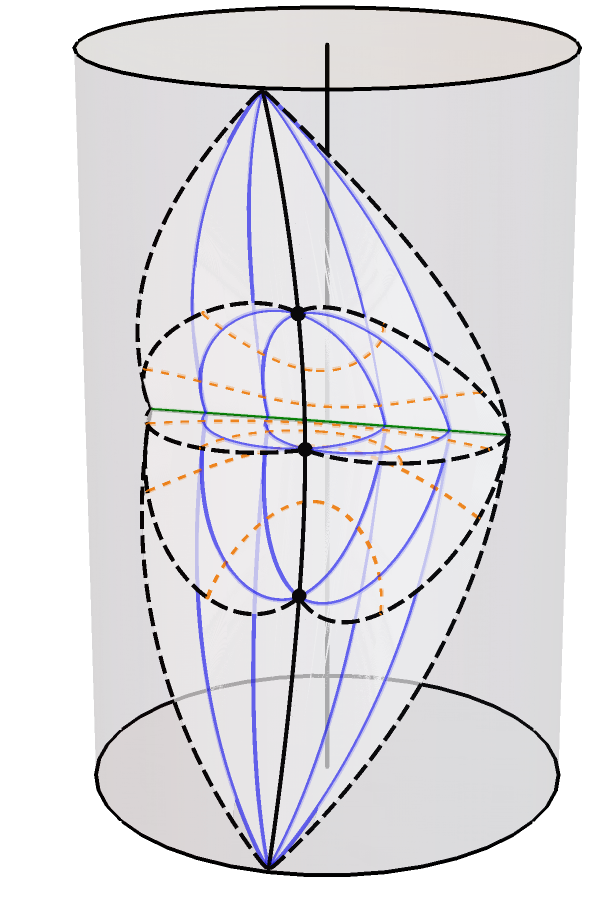}
  \hfill
  \includegraphics[width=0.45\textwidth]{%
    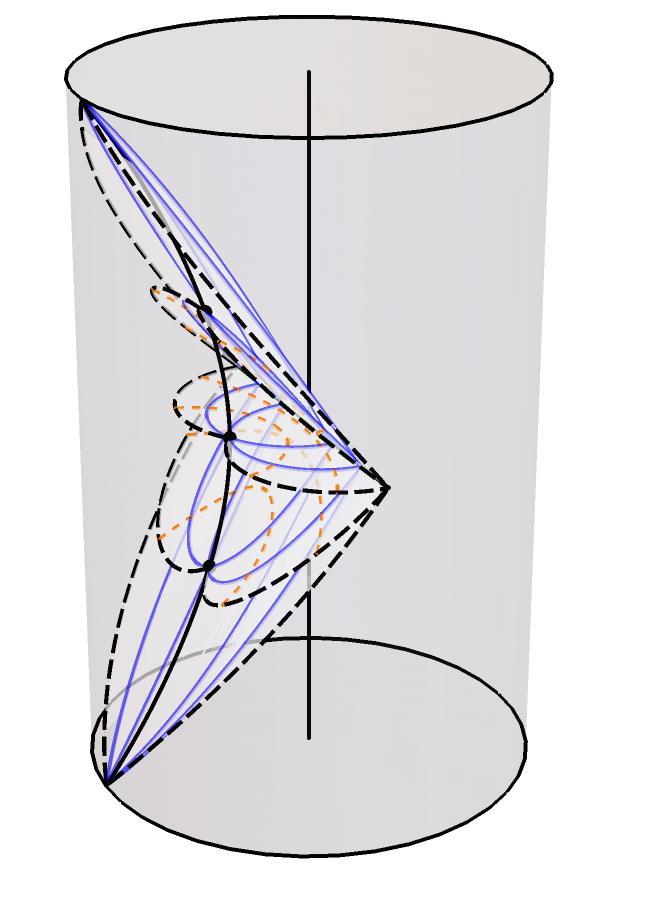}
  \caption{
      The Class I\textsubscript{rapid,B} solution
      embedded within global AdS$_3$.
      The particle worldline is shown in solid black.
      Several surfaces of constant $t$ are plotted.
      The event horizons are demonstrated by the surfaces at early and late $t$.
      The bifurcation surface is shown as a green line.
      The surfaces $x=x_+$ and $x'=x_+$ identified.
      Lines of constant $t$ within these surfaces
      are shown in long-dashed black.
      None of the conformal boundary is included in the solution.
      Lines of constant $x$ are shown in blue,
      with lines of constant $y$ in dashed orange.
      To guide the eye, the axis of the cylinder is also shown in black.}
  \label{fig:3DIstringrapidheavy}
\end{figure}

Further, within the rapidly accelerating and saturated particle solutions, 
a new phenomenon now occurs that has no analogy in the 
4-dimensional case. Due to the polar origin ($r=0$ or $y\to-\infty$)
being displaced from the centre of AdS by acceleration, the lines of constant
$x$ or $\phi$ now arc through the bulk, and only reach the boundary if
$y\to x$ is allowed on that curve. For slowly accelerating solutions,
all values of $y$ are allowed and this always occurs, but for $A\ell>1$, 
$y$ is now constrained by the presence of the horizon, and if $x_+> -y_h$, 
then no constant $x$ line can reach the boundary, hence the geometry 
has a finite spatial section. In effect, the point particle has become 
sufficiently ``heavy'' that it has cut out all of the conformal boundary.
This is shown in figure \ref{fig:3DIstringrapidheavy}.
In this phase, the bifurcation surface compactifies to become a circle
and the horizon structure is similar to that of de Sitter space.

\subsubsection{The holographic mass}
In this subsection, we identify the holographic mass
of the slowly accelerating conical deficit.

Following the prescription of Brown and York \cite{Brown:1992br},
a quasi-local gravitational energy can be obtained 
by varying the action with respect to the boundary metric,
provided the variational problem is well-posed.
To correctly formulate the variational problem 
for an asymptotically AdS spacetime,
one can add counterterms to regulate the
UV divergences arising near the boundary
\cite{Balasubramanian:1999re,Mann:1999pc}. 

Following the recipe,
one expands the metric in the Fefferman-Graham frame  
\begin{equation}
  ds^2=
  - \frac{\ell^2}{z^2}dz^2
  + \frac{\ell^2}{z^2}
  \left( \g0 +z^2\g2 +z^4\g4 \right)
  \,.
  \label{FGexpansion}
\end{equation}
Here, $\g0$, $\g2$, and $\g4$ are covariant two-tensors.
Note that this expression is exact;
in $2+1$ dimensions the series terminates at order $z^2$ 
\cite{Henningson:1998gx,Balasubramanian:1999re,deHaro:2000vlm}. 
The holographic stress tensor $\left\langle T \right\rangle$
is completely determined by $\g0$ and $\g2$:
\begin{equation}
  \left\langle T \right\rangle =
  -\frac{\ell}{8\pi}
  \left(\g2-\g0\Tr[\g0^{-1}\g2]\right)
  \,.
  \label{eq:stressTensor}
\end{equation}

We now transform the metric \eqref{eq:metricxy2}
to Fefferman-Graham gauge near the boundary,
as in \cite{Anabalon:2018ydc}, using the transformation
\begin{equation}
  y =
  \xi
  + \sum\limits_{m=1}^{\infty}
  F_m(\xi) \left(\frac{z}{\ell}\right)^m
  \,,
  \qquad
  x =
  \xi
  + \sum\limits_{m=1}^{\infty}
  G_m(\xi) \left(\frac{z}{\ell}\right)^m
  \,.
\label{eq:expan}
\end{equation}
It is then straightforward to solve for the unknown functions order by order. 
For example, to determine $G_1$ we perform the above transformation,
expand the resulting metric to order $z^{-2}$,
and enforce the lack of a cross term $g_{z\xi}=0$.
We then expand to order $z^{-1}$, then $z^{-2}$, and so on,
sequentially fixing the $F_m$'s and $G_m$'s to acquire Fefferman-Graham form.
In this process, $F_1$ remains unfixed,
appearing as a conformal factor in the boundary metric $\g0$.
Also, the metric expansion explicitly terminates at $\mathcal{O}(z^2)$:
once one gauge fixes $g_{zz}=\ell^2z^{-2}$ and $g_{z\xi}=0$ at order $m=7$, 
the other metric components at the same order vanish identically. 
This property persists at orders $m>7$;
one is always able to add terms to the coordinate expansions \eqref{eq:expan} 
such that the $\g{n}$ with $i\geq 5$ are identically zero to arbitrary order.

For notational convenience, we define a dimensionless function
$\Upsilon(\xi)\equiv 1-A^2\ell^2(1-\xi^2)$, then a
strategic redefinition of the conformal factor
\begin{equation}
  F_1(\xi) = \frac{\Upsilon^{3/2}}{ A\ell \omega(\xi) }
  \label{eq:Iconformalfactor}
\end{equation}
in terms of a new (non-zero) function $\omega$ 
simplifies the boundary metric found by the process described above:
\begin{equation}
  \g0 = 
  \frac{\omega(\xi)^2}{A^2}
  \left[
    d\tau^2 
    - A^2\ell^2 \frac{d\xi^2}{{1-\xi^2}}
  \right]
  \,.
  \label{eq:Ig0}
\end{equation}
Note that $\g0$ has the correct dimensions of length squared.
The leading correction is also diagonal:
\begin{multline}
  \g2 = 
  \frac{1}{2 A^2\ell^2}
  \left[
    1-A^2\ell^2 
    +(1-\xi^2)\Upsilon^2
    \left(\frac{\omega'(\xi)}{\omega(\xi)}\right)^2
  \right] d\tau^2
  + \Bigg[
    \frac{1-A^2\ell^2}{2(1-\xi^2)\Upsilon^2}
    \\
    + \left(1-3A^2\ell^2(1-\xi^2)\right)
    \frac{\xi}{(1-\xi^2)\Upsilon}
    \left(\frac{\omega'(\xi)}{\omega(\xi)}\right)^2
    +\frac{3}{2}\left(\frac{\omega'(\xi)}{\omega(\xi)}\right)^2
    -\frac{\omega''(\xi)}{\omega(\xi)}
  \Bigg] d\xi^2
  \,.
  \label{eq:Ig2}
\end{multline}
Note that $\g2$ is dimensionless, as expected.
We will not write down $\g4$ as it will not be needed.
Raising and lowering indices with $\g0$, 
the non-zero components of the stress tensor are
\begin{multline}
  16\pi \omega(\xi)^2
  \ell
  \langle T^{\tau}_{\tau} \rangle 
  =
  -\left(1-A^2\ell^2\right)
  - 2\xi\left(1-3A^2\ell^2\left(1-\xi^2\right)\right)
  \Upsilon
  \left(\frac{\omega'(\xi)}{\omega(\xi)}\right)
  \\
  + \left(1-\xi^2\right)\Upsilon^2
  \left[
      2\frac{\omega''(\xi)}{\omega(\xi)}
    - 3\left(\frac{\omega'(\xi)}{\omega(\xi)}\right)^2
  \right]
  \label{eq:Istresstautau}
\end{multline}
and
\begin{equation}
  16\pi \omega(\xi)^2
  \ell
  \langle T^{\xi}_{\xi} \rangle
  =
  \left(1-A^2\ell^2\right)
  + \left(1-\xi^2\right)
  \Upsilon^2
  \left(\frac{\omega'(\xi)}{\omega(\xi)}\right)^2
  \,.
  \label{eq:Istressxixi}
\end{equation}
As expected for a two-dimensional boundary theory
\cite{Brown:1986nw, Henningson:1998gx}, 
tracing this stress tensor reproduces the usual Weyl anomaly 
\begin{equation}
    \langle \Tr[\g0^{-1} T] \rangle =
    \frac{c_{\text{Virasoro}}}{24\pi}R(\g0)
    \,,
    \label{eq:anomaly}
\end{equation}
where $R(\g0)$ is the Ricci scalar of the boundary metric $\g0$:
\begin{multline}
  R(\g0) =
  -\frac{2\Upsilon}{\ell^2\omega(\xi)^2}
  \Bigg(
    \xi\left(1-3A^2\ell^2(1-\xi^2)\right)\frac{\omega'(\xi)}{\omega(\xi)}
    \\
    +(1-\xi^2)\Upsilon
    \left[
      \left(\frac{\omega'(\xi)}{\omega(\xi)}\right)^2
      -\frac{\omega''(\xi)}{\omega(\xi)}
    \right]
  \Bigg)
  \,.
  \label{eq:Iriccig0}
\end{multline}
The central charge is calculated to be $c_{\text{Virasoro}}=3\ell/2$,
suggesting the existence of a locally Virasoro asymptotic symmetry group.
The stress tensor is covariantly conserved: $\nabla_\mu {T^\mu}_\nu=0$.

\begin{figure}[t!]
\centering
\includegraphics[width=0.5\textwidth]{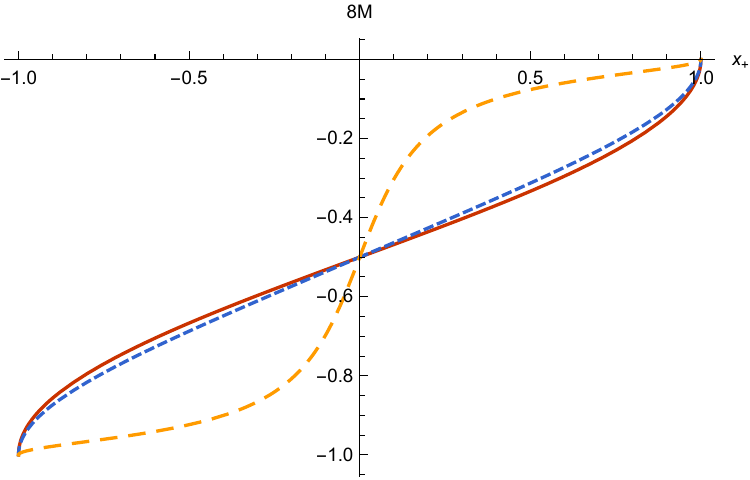}~
\includegraphics[width=0.5\textwidth]{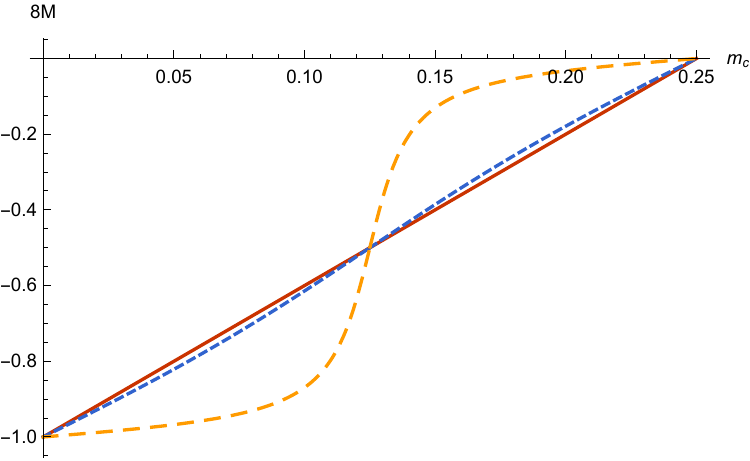}
\caption{The holographic mass of the conical deficit pulled by a wall
plotted as a function of $x_+$ (Left) and the local ``conical mass'' $m_c$ (Right)
with various acceleration parameters: $A=0$ (solid red),
$A=0.5\ell$ (dotted blue), $A=0.99\ell$ (dashed orange).  }
\label{fig:Imassplots}
\end{figure}

We are now in a position to write down the mass.
Being careful to include the contributions from both $(x,y)$ patches,
accounting for their orientations, the mass,
calculated with respect to $\partial_t$, is
\begin{equation}
  M =
  2A\int_{x_+}^{1}
  \sqrt{-\det\g0}
  \left\langle T_\tau^\tau\right\rangle
  d\xi
  \,,
  \label{eq:classImassIntegral}
\end{equation}
where the integrand is
\begin{multline}
  \sqrt{-\det\g0} \left\langle T^\tau_\tau \right\rangle
  =
  -
  \frac{1}{8\pi A \alpha \sqrt{1-\xi^2}}
  \Bigg\{
    \frac{1-A^2\ell^2}{2\Upsilon}
    + \xi\left(1-3A^2\ell^2(1-\xi^2)\right)
    \left(\frac{\omega'(\xi)}{\omega(\xi)}\right)
    \\
    +(1-\xi^2)\Upsilon
    \left[
      \frac{3}{2}\left(\frac{\omega'(\xi)}{\omega(\xi)}\right)^2
      -\frac{\omega''(\xi)}{\omega(\xi)}
    \right]
  \Bigg\}
  \,.
  \label{eq:classImassIntegrand}
\end{multline}
The dependence upon the conformal representative arises 
as a result of the usual transformation properties 
of the stress tensor for a two-dimensional CFT.
The stress tensor in two dimensions is not a primary operator
(it is only quasi-primary).
Such an operator transforms under conformal mappings
with additional terms involving a Schwarzian derivative \cite{Blumenhagen:2009zz}.
Since in dimensions greater than two all quasi-primary fields are primary,
such terms do not appear in the boundary stress tensor 
for the four-dimensional C-metric.
In order to obtain the correct holographic information in two dimensions,
we need to choose a particular conformal representative of the boundary metric.
For three-dimensional gravity, we should choose the Brown-Henneaux condition 
\cite{Brown:1986nw,Banados:1998gg, Banados:1998sm},
setting the boundary metric to be a cylinder.
In fact, this is necessary for a well defined variational problem 
\cite{Papadimitriou:2005ii}.
This choice corresponds to selecting the ground state for the boundary CFT.
We thus fix the conformal factor to be $\omega=1$.
The integral then takes the form
\begin{equation}
M = -
    \frac{1}{8\pi \alpha}
    \sqrt{1-A^2\ell^2}
    \int_{x_+}^{1}
    \frac{d\xi}{\Upsilon\sqrt{1-\xi^2}}
    \,,
\end{equation}
which is readily integrated, yielding
\begin{equation}
  M =
    -
    \frac{1}{8\pi \alpha}
    \sqrt{1-A^2\ell^2}
    \left(
    \frac{\pi}{2}
    - \arctan\left[\frac{x_+}{\sqrt{1-A^2\ell^2}\sqrt{1-x_+^2}}\right]
    \right)
    \,.
\end{equation}
To make this expression more intuitive,
let's use the parameter $K$ instead
and substitute the value of $\alpha$ from \eqref{eq:Ialpha}.
The mass is then written
\begin{equation}
  M =
  -
  \frac{1}{8\pi}
  \left(
  \frac{\pi}{2}
  - \arctan\left[
      \frac{\cot\left(\frac{\pi}{K}\right)}{\sqrt{1-A^2\ell^2}}
    \right]
  \right)
  \,.
  \label{eq:Istringmassk}
\end{equation}
The mass is plotted for various acceleration parameters
in figure \ref{fig:Imassplots}.
This shows the effect of acceleration as introducing a discrepancy between
the holographic mass and the local conical mass. For no acceleration, there is
a simple linear relation between the two, as one would expect. For even relatively
significant acceleration ($A=0.5\ell$) this situation is not much altered. However,
as the acceleration approaches the slow limit $A\to\ell$, we see a marked
effect on the holographic mass. At first, there is very little alteration in $M$
as $m_c$ is increased, then around a deficit angle of $\pi$, there is a sharp 
and steep crossover to very small $M$, with once again very little change as
the deficit angle gradually closes off the spacetime.

As a check, one can take the limit of vanishing acceleration parameter $A$.
We recover exactly the Casimir energy of pure (global) AdS$_3$
with a conical deficit
\cite{Balasubramanian:1999re,Henningson:1998gx,Carlip:1995zj}
\begin{equation}
  M\textsubscript{AdS$_3$} = - \frac{1}{8K}
  \,.
  \label{eq:globalAdS3mass}
\end{equation}

\subsection{A particle pushed by a strut}
\label{sec:Istrut}

We now just briefly comment on the counterpart solution of a particle being
pushed by a strut. In fact, the asymptotically flat counterpart of this construction
(which can be achieved by taking the $\ell\rightarrow\infty$ limit) 
has been studied by Anber \cite{Anber:2008zz}.
This solution shares many features in common 
with the solution of section \ref{sec:Istring}
so we will discuss it in less detail.

As before, we use the polar representation of the geometry, which is almost 
in the form of \eqref{eq:classIrphi}, but instead set $x=-\cos(\phi/K)$
with the parameter $K = \pi/\arccos\left(-x_+\right)$
The metric then takes the form
\be
\beal
ds^2 &= \frac{1}{\left[1-Ar\cos\left(\phi/K\right)\right]^2}
\left[ f(r)\frac{dt^2}{\alpha^2} -\frac{dr^2}{f(r)}
-r^2 \frac{d\phi^2}{K^2} \right] \,, \\
f(r) &= 1 + (1-A^2\ell^2) \frac{r^2}{\ell^2} \,.
\eeal
\label{eq:classIstrutrphi}
\ee
By similar arguments to those of section \ref{sec:Istring},
we find that one should take $\alpha=\sqrt{\left\vert 1-A^2\ell^2 \right\vert}$.
The local acceleration of the particle is again
$\left\vert\mathbf{a}\right\vert = A$, and the particle mass and wall 
tension are, respectively, 
\begin{equation}
m_c=\frac{1}{4}\left(1-\frac{1}{K}\right)
\end{equation}
and
\begin{equation}
\sigma = -\frac{A}{4\pi}\sin\left(\frac{\pi}{K}\right)  \,.
\label{eq:tensionIslowStrut}
\end{equation}

By mapping the solution to a portion of global space, we obtain figure 
\ref{fig:3DIstrutslow} for the slowly accelerating deficit.
The details of the mapping are given in appendix \ref{app:map}.
The interpretation of this figure is similar 
to that for the particle accelerated by a wall,
(cf. figure \ref{fig:3DIstringslow}
and the discussion given in section \ref{sec:Istring}).
\begin{figure}[t!]
\centering
\includegraphics[width=0.5\textwidth]{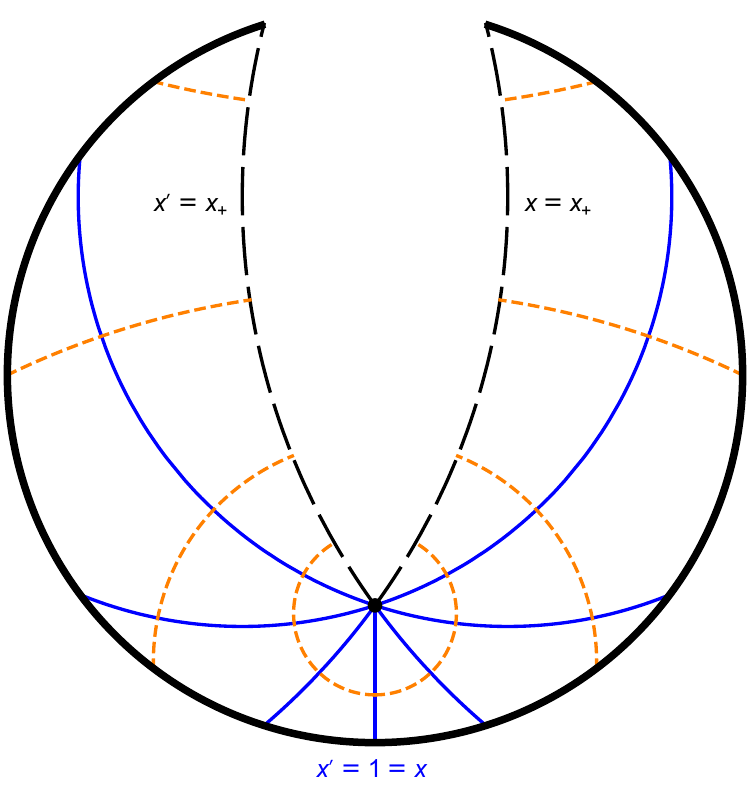}
\caption[ The slowly accelerating conical deficit, pushed by a strut]
{The slowly accelerating conical deficit with $A=0.9\ell$,
pushed by a strut, mapped onto the Poincar\'{e} disk.
The deficit is shown as a black point and accellerates southwards.
A ``wedge'' is removed from the global space, with its edges
$x=x_+$ and $x'=x_+$
(long-dashed black) identified to attain a wall.
Several lines of constant $x$ are shown in blue,
with lines of constant $y$ in dashed orange.}
\label{fig:3DIstrutslow}
\end{figure}

\begin{figure}[t!]
\centering
\includegraphics[width=0.45\textwidth]{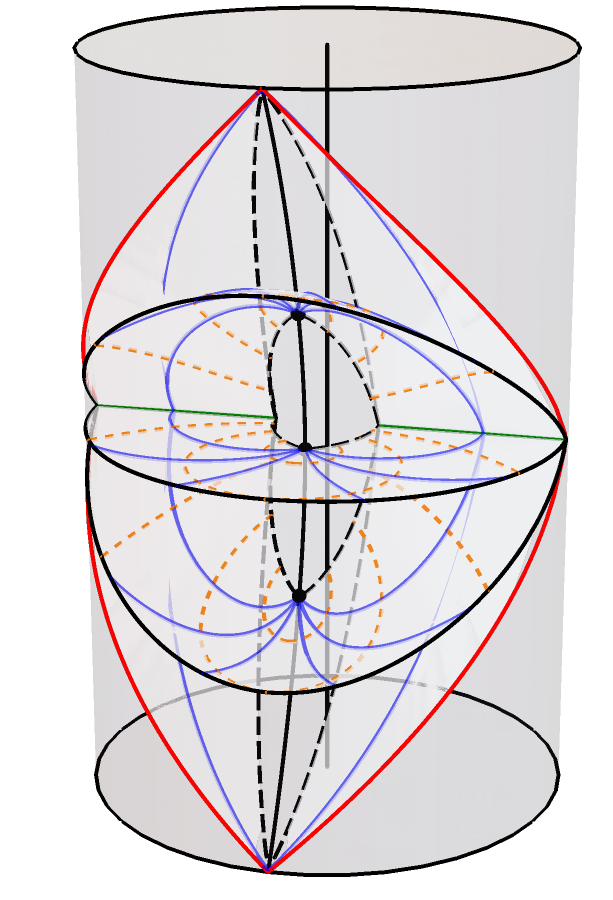}
\hfill
\includegraphics[width=0.45\textwidth]{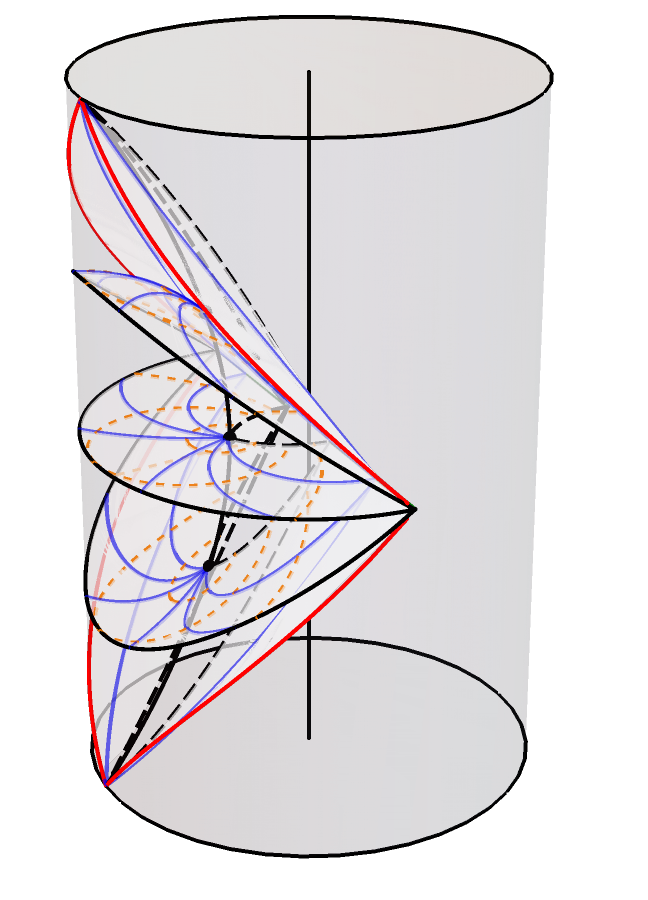}
\caption{The rapidly accelerating light conical deficit with $A>\ell^{-1}$,
pushed by a strut, embedded within global AdS$_3$.
The deficit's worldline is shown in solid black.
Several surfaces of constant $t$ are plotted.
The event horizon is demonstrated by the surfaces at early and late $t$.
The bifurcation surface is shown as a green line.
A ``wedge'' is removed, with its faces $x=x_+$ and $x'=x_+$ identified.
These faces are indicated by long-dashed black curves
within each time-slice,
and in dashed red at the boundary.
The boundary of the classically accessible
subset of the global boundary is shown in solid red.
Lines of constant $x$ are shown in blue,
with lines of constant $y$ in dashed orange.
To guide the eye, the axis of the cylinder is also shown in black.}
\label{fig:3DIstrutrapid}
\end{figure}

In the rapid and saturated phases, much like with the particle accelerated by a wall,
changing the magnitude of the conical deficit can induce a phase transition
in the global structure of the spacetime.
For light conical deficits, the gluing surface $x=x_+$
connects the conical deficit's worldline to a horizon (in the patch 
covered by \eqref{eq:classIstrutrphi}.
One can also view this as removing a ``wedge'' which includes part of the 
acceleration horizon from the Rindler wedge.
This situation arises for the I\textsubscript{rapid,B} and I\textsubscript{saturated,D}
single-strut solutions and the former is shown in figure
\ref{fig:3DIstrutrapid}.

\begin{figure}[t!]
\centering
\includegraphics[width=0.45\textwidth]{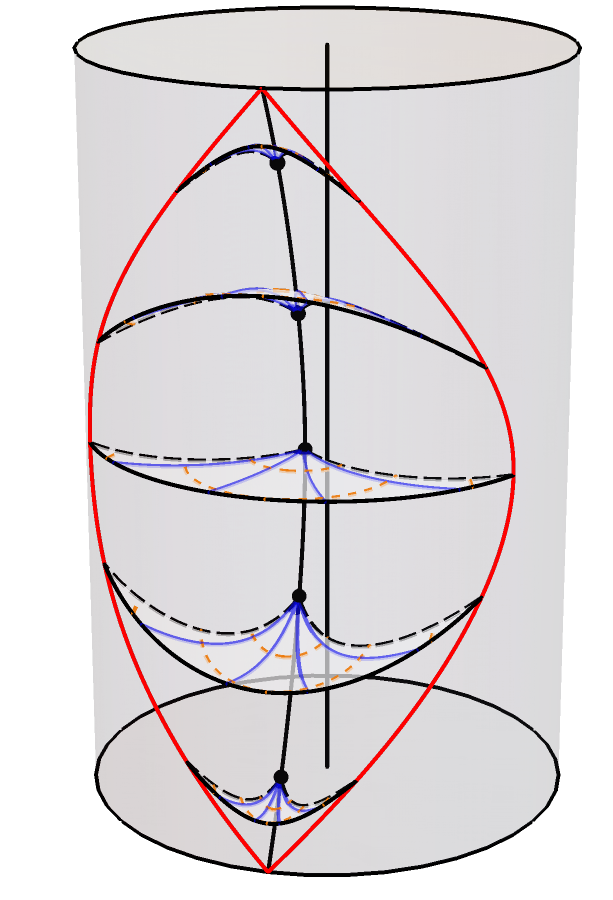}
\hfill
\includegraphics[width=0.45\textwidth]{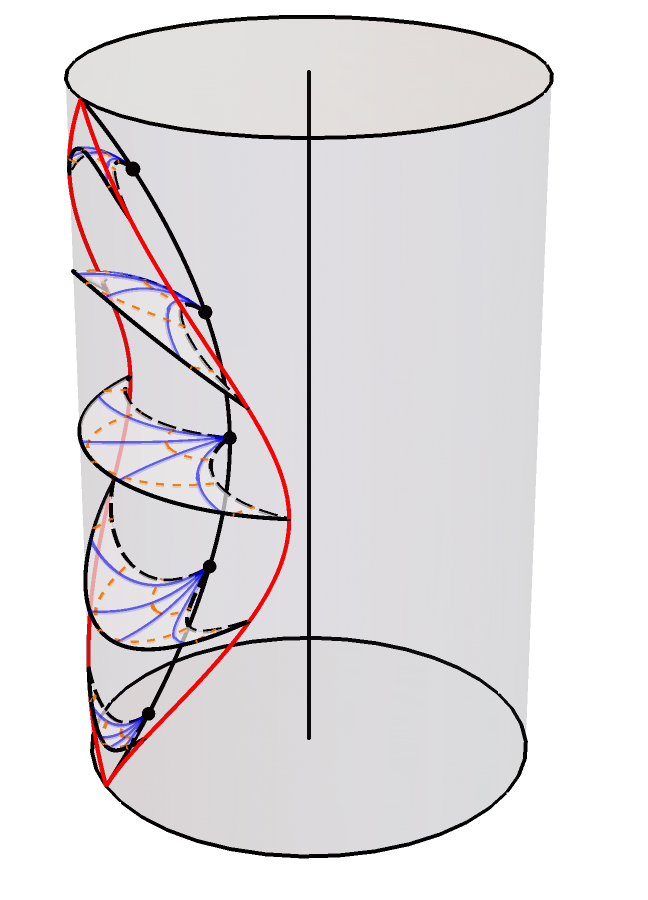}
\caption[The Class I\textsubscript{rapid,B} solution
- a heavy, rapidly accelerating conical deficit]
{The rapidly accelerating heavy conical deficit with $A>\ell^{-1}$,
pushed by a strut, embedded within global AdS$_3$.
The deficit's worldline is shown in solid black.
Several surfaces of constant $t$ are plotted.
The surfaces $x=x_+$ and $x'=x_+$ are identified.
Edges of $x=x_+$ are shown in long-dashed black within each time-slice.
$x_+$ is taken positive enough that the entirety of the boundary
is removed from the space before identification.
The boundary of the classically accessible
subset of the global boundary is shown in red.
Lines of constant $x$ are shown in blue,
with lines of constant $y$ in dashed orange.
To guide the eye, the axis of the cylinder is also shown in black.}
\label{fig:3DIstrutrapidheavy}
\end{figure}

On the other hand, for heavier conical deficits, the identified surfaces 
connect the deficit's worldline to the conformal boundary. This situation 
arises for the I\textsubscript{rapid,A} and I\textsubscript{saturated,C}
single-strut solutions. The portion of space then removed from the Rindler 
geometry is then so large that it includes the entirety of the horizon, and we 
are left with a small segment of spacetime extending from the particle to the 
boundary as shown in figure \ref{fig:3DIstrutrapidheavy}.

\subsubsection{The holographic mass}
Considering again the slowly accelerating phase, the holographic calculation 
proceeds identically to that of section \ref{sec:Istring}.
In particular, the expressions for the Fefferman-Graham expansion
(\ref{eq:Iconformalfactor}, \ref{eq:Ig0}, \ref{eq:Ig2}),
the stress tensor (\ref{eq:Istresstautau}, \ref{eq:Istressxixi}),
and the Ricci scalar of the boundary metric \eqref{eq:Iriccig0} all hold.

The limits for the mass integration must be updated,
\begin{equation}
M = 2A\int_{-1}^{x_+} \sqrt{-\g0}\langle T^{\tau}_{\tau} \rangle d\xi \,,
\end{equation}
however evaluation gives the same result \eqref{eq:Istringmassk}.
Thus, it appears that although one might na\"{i}vely 
associate a negative mass with the negative tension domain wall 
\cite{Anber:2008zz}, this is erroneous.
The mass again reduces to that of global AdS$_3$ with a conical deficit
in the limit of vanishing $A$.

\section{Accelerating BTZ black holes}
\label{sec:accBTZ}

The BTZ black hole is formed by taking a different identification of AdS.
First, one transforms to a Rindler wedge (or planar BTZ) patch of AdS,
then identifies along constant $x$-lines. In the absence of acceleration
(taking the limit $A\to0$ of \eqref{eq:metricxy2} with judicious coordinate
rescalings) these $x=$const.\ lines have vanishing extrinsic curvature. 
Once we have finite $A$ however, the way in which the coordinates slice the 
Rindler wedge is skewed (see figures \ref{fig:3DIIstrutslow} \& \ref{fig:3DIIstrutrapid}), 
so that even though the patch
of AdS is the same as for planar BTZ, the new constant $x$ surfaces now
have non-vanishing extrinsic curvature. While we have long been aware that
AdS can be sliced in many ways, this new phenomenon has not been noticed
in the literature to our knowledge.

\subsection{A BTZ black hole pushed by a strut (Class II\textsubscript{right})}
\label{sec:IIstrut}

Starting from a patch of Class II spacetime \eqref{eq:metricxy2} with $x>1$, 
as presented in table \ref{tab:3},
one may construct a one parameter extension of the static BTZ solution
describing a black hole with a strut emerging from its horizon.
The existence of such a black hole was proposed in \cite{Astorino:2011mw},
although several features,
including the possibility of a ``rapid'' phase
possessing a non-compact acceleration horizon,
went unacknowledged.
We seek to clarify the discussion and highlight these additional features here.
\begin{figure}[tb!]
\centering
\includegraphics[width=0.49\linewidth]{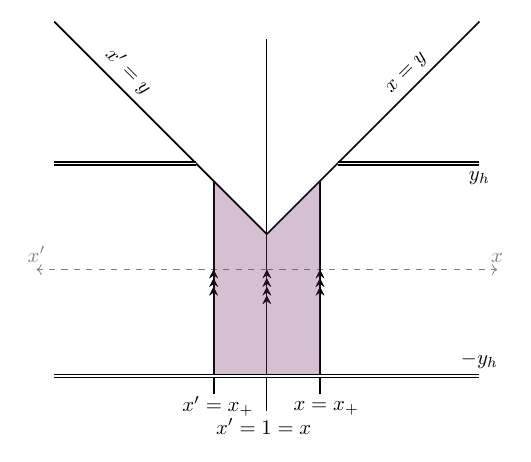}
\includegraphics[width=0.49\linewidth]{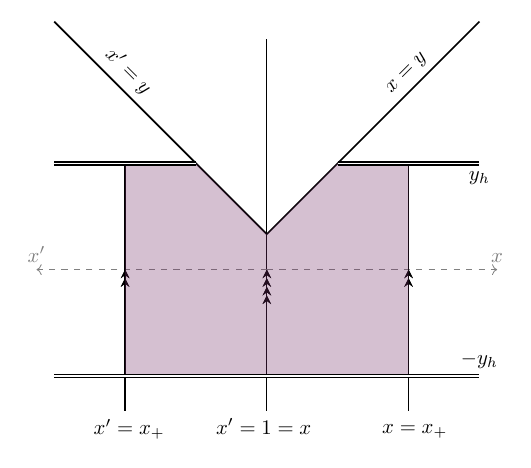}
\caption{{\bf Left:} II\textsubscript{right,slow}. 
The two patches of type II spacetime -- with $x>1$ -- and the identifications used 
to construct a black hole accelerated by a pushing strut when $x_+<y_h$.
{\bf Right:} II\textsubscript{right,rapid}. The two patches of type II spacetime -- 
with $x>1$ -- and the identifications used to construct a black hole accelerated 
by a pushing strut when $x_+>y_h$.}
\label{fig:IIrightslowrapid}
\end{figure}%

Choose some value $x_+>1$
and define a patch with $x\in[1,x_+)$.
Glue two copies of this patch, mirroring along both $x=1$ and $x=x_+$.
The identifications are shown in figure \ref{fig:IIrightslowrapid}.
The $x=1$ axis of the newly formed spacetime is regular,
while $x=x_+$ marks the position of a domain wall
of negative tension $\sigma=-A(4\pi )^{-1}\sqrt{Q(x_+)}$.
The resulting spacetime describes the exterior of a black hole 
with horizon located at $-y_{h}=-\sqrt{1+A^{-2}\ell^{-2}}$.
If $x_+$ is larger than $y_h$, 
there is also a ``droplet'' horizon \cite{Hubeny:2009ru}
present at $y=y_h$ for $x>y_h$.
This rapid phase is a novel feature unnoticed in \cite{Astorino:2011mw}.
The existence of such a phase strengthens the analogy
with the four-dimensional accelerating black hole.

To cast the metric in more intuitive coordinates,
take $x=\cosh\left(\psi/K\right)$ where $K=\pi/\arcosh(x_+)$.
Defining dimensionful coordinates $\rho=-(Ay)^{-1}$ and $t=\alpha A^{-1}\tau $,
the metric becomes
\begin{equation}
\begin{split}
  ds^2 &=
    \frac{1}{\left(1+A \rho \cosh\left(\frac{\psi}{K}\right)\right)^2}
    \left(
    f(\rho) \frac{dt^2}{\alpha^2}
    - \frac{d\rho^2}{f(\rho)}
    - \rho^2 \frac{d\psi^2}{K^2}\right)
  \,,
  \\
  f(r) &=
    -1+(1+A^2\ell^2)\rho^2/\ell^2
  \,,
\end{split}
  \label{eq:IIrightmetricrhosig}
\end{equation}
where again we leave room for the possibility that
$t$ should be scaled by some dimensionless quantity $\alpha$.
The parameter $K$ has been chosen such that
the range of $\psi$ is $(-\pi,\pi)$.
We require both $K>0$ and $\pi/K<\arcosh(y_h)$.
This patch does not cover the entire region between 
the black hole horizon and the boundary
(and the acceleration horizon).
The region immediately exterior to the black hole has $r>0$.
Proceeding from the black hole along a line of constant $\psi$
in the direction of increasing $r$, one encounters a 
coordinate singularity as $r\rightarrow\infty$, 
which is a geometrically uninteresting locus.
One must take a second patch with $r<0$ to cover the region
bounded by the conformal boundary (for small $\psi$)
and possibly also the acceleration horizon (for large $\psi$).
This is exactly analogous to the slowly accelerating black hole
in 4D.
The conformal boundary is then approached as
$r\rightarrow-\left[A\cosh\left(\psi/K\right)\right]^{-1}$.
There is a regular semi-axis at $\psi=0$,
while along $\psi=\pm\pi$ there lies a domain wall of (negative) tension
\begin{equation}
  \sigma = - \frac{A}{4\pi } \sinh\left(\frac{\pi}{K}\right)
  \,.
\end{equation}

These Class II\textsubscript{right} solutions form a one parameter extension 
of the well-known family of static BTZ black holes.
To highlight the relationship, take the metric \eqref{eq:IIrightmetricrhosig} 
and make the parameter redefinitions
$K=m^{-1}$ and $A=m\mathcal{A}$.
Also, make the coordinate rescalings $r=m\rho$ and $\tilde{t} = m^{-1}t$.
The metric becomes
\begin{equation}
\begin{split}
  ds^2 &= 
  \frac{1}{\Omega(r,\psi)^2}
  \left[
    F(r) \frac{d\tilde{t}^2}{\alpha^2}
    - \frac{dr^2}{F(r)}
    - r^2 d\psi^2
  \right]
  \,,
  \\
  F(r) &= -m^2(1-\mathcal{A}^2r^2) + \frac{r^2}{\ell^2}
  \,,
  \\
  \Omega(r,\psi) &= 1 + \mathcal{A} r \cosh(m\psi) 
  \,.
\end{split}
  \label{eq:ClassIIrightmetcal}
\end{equation}
We require that both $m>0$ and the slow acceleration condition -- 
which is now $m\sinh\left(m\pi\right)<\left(\mathcal{A}\ell\right)^{-1}$
-- hold. A chart with $r>0$ covers the region immediately exterior to
the black hole horizon, which lies at $r=m\ell(1+m^2\mathcal{A}^2\ell^2)^{-1/2}$.
A second chart with $r<0$ covers the region bordered by the conformal boundary,
which lies at $r\rightarrow -\left[\mathcal{A}\cosh(m\psi)\right]^{-1}$.

\begin{figure}[tb!]
  \centering
  \includegraphics[width=0.45\textwidth]{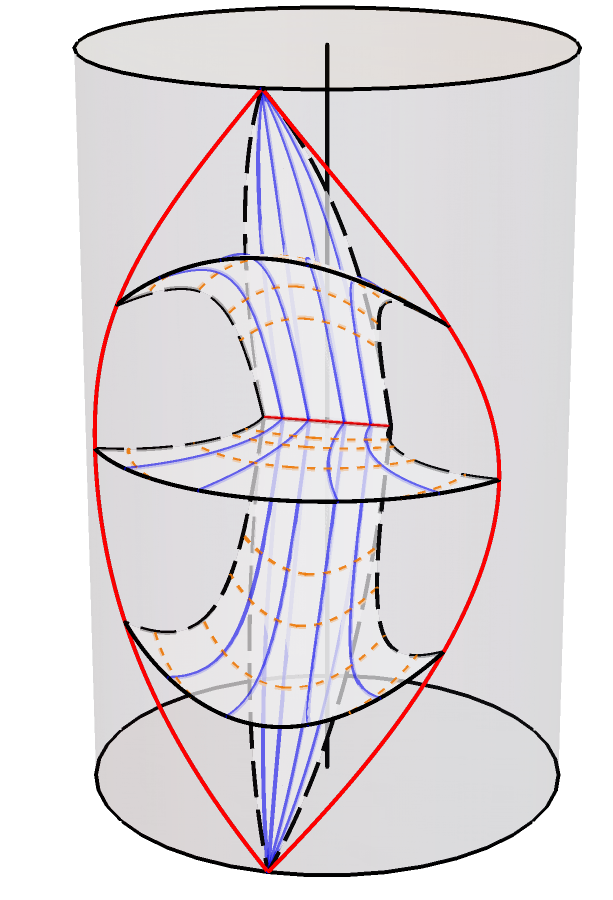}
  \hfill
  \includegraphics[width=0.45\textwidth]{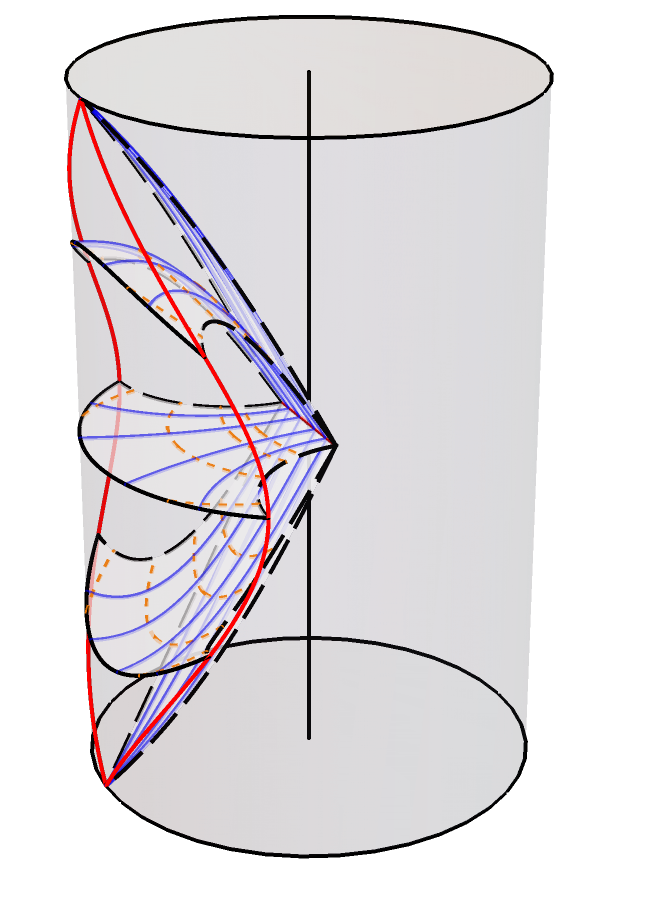}
  \caption{ 
    The Class II\textsubscript{right, slow} black hole.
    Several surfaces of constant $t$ are shown.
    The lines $x=x_+<y_h$ are shown in dashed black,
    and are identified with their partner within the same time slice,
    wrapping the bifurcation surface (shown as a horizontal red line)
    into a circle.
    The early and late time-slices demonstrate the event horizon.
    Lines of constant $y$ are shown in dashed orange.
    Lines of constant $x$ are shown in blue.
    The classically accessible region of the boundary is delimited in red.
  }
  \label{fig:3DIIstrutslow}
\end{figure}

\begin{figure}[tb!]
  \centering
  \includegraphics[width=0.45\textwidth]{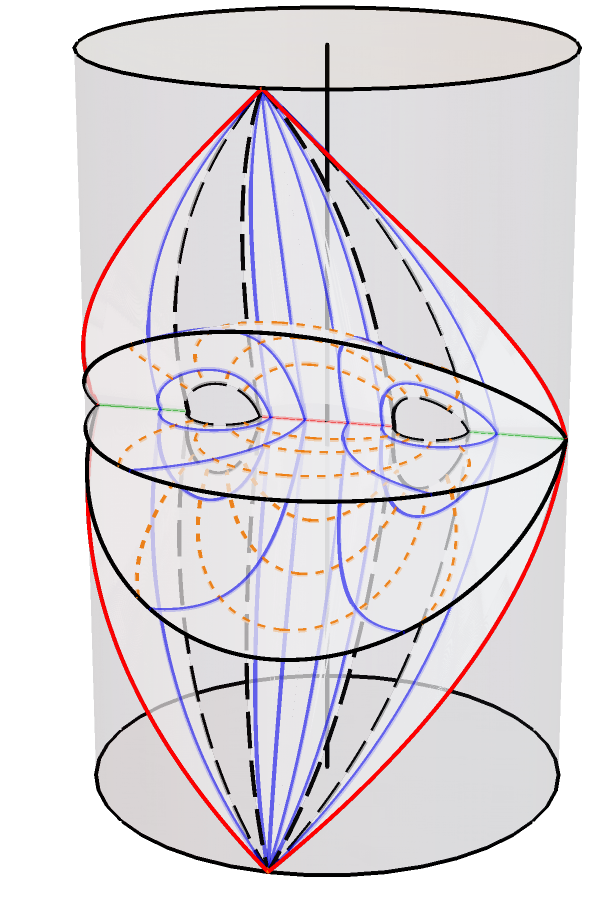}
  \hfill
  \includegraphics[width=0.45\textwidth]{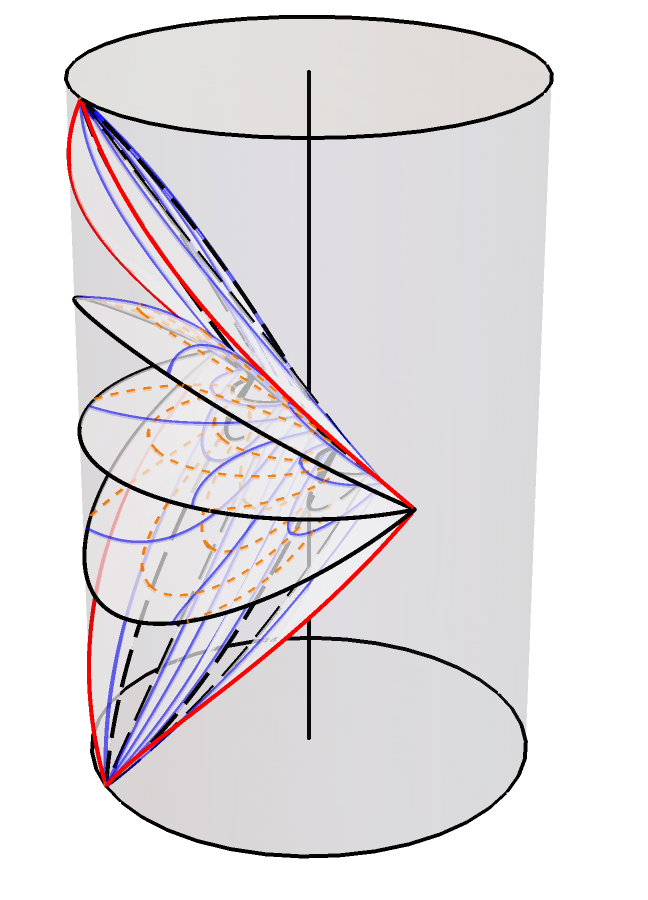}
  \caption{
    The Class II\textsubscript{right,rapid} black hole.
    Several surfaces of constant $t$ are shown.
    The lines $x=x_+<y_h$ are shown in dashed black,
    and are identified with their partner within the same time slice.
    The early and late time-slices show the event horizons.
    The black hole event horizon has a compact bifurcation surface
    with topology $S^1$, shown in red,
    while the green line denotes the ``acceleration horizon'' or droplet's
    bifurcation line with topology $\mathbb{R}$.
    Lines of constant $y$ are shown in dashed orange.
    Lines of constant $x$ are shown in blue.
    The classically accessible region of the boundary is delimited in red.
  }
  \label{fig:3DIIstrutrapid}
\end{figure}

To understand the solution, we can again perform a mapping to global AdS$_3$,
the details of which are given in appendix \ref{app:map}.
A solution in the slow phase $x_+<y_h$
is shown in figure \ref{fig:3DIIstrutslow},
while one in the rapid phase $x_+>y_h$
is shown in figure \ref{fig:3DIIstrutrapid}.
In the slow phase, there is a single bifurcate Killing horizon generated by
$\partial_t$.
Upon making the appropriate identification,
the bifurcation surface of this horizon acquires the topology of a circle.
All of the lines of constant $x$ (and $t$) connect the bifurcation surface 
of this horizon to the conformal boundary (consider the blue lines in 
figure \ref{fig:3DIIstrutslow}).
This construction mirrors that of the usual, static BTZ black hole 
(cf.\ figure \ref{fig:3Dbtz}).
Of course, when constructing the Class II\textsubscript{right,slow} black hole,
surfaces of constant $x$ are glued resulting in a strut.
In the usual BTZ construction the surfaces to identify are specifically chosen
by quotienting (the universal covering space of) AdS$_3$
by the group of integers to maintain regularity.
In the rapid phase, $\partial_t$ generates two disjoint bifurcate horizons.
There exist lines of constant $x>y_h$ (and $t$)
which connect the black hole bifurcation surface to a second, disjoint
bifurcation surface with topology $\mathbb{R}$.
While the slow phase is qualitatively similar
to the traditional, static BTZ solution, albeit
with induced tension in the identified surface, the
rapid phase hitherto absent from the literature is qualitively distinct.

\subsubsection{The holographic mass}
\label{subsec:IIstrutMass}
%
Proceeding as for the accelerating deficits,
we calculate the holographic mass from the metric \eqref{eq:metricxy2}
with the Class II metric functions given in table \ref{tab:3}.
Once again, it is difficult to define the mass 
of a gravitational solution in the presence of a non-compact horizon.
Sidestepping this issue,
we focus on the case where the parameter $A$
is small enough that
\begin{equation}
  \pi/K<\arcosh\left(\sqrt{1+A^{-2}\ell^{-2}}\right)
  \,.
\end{equation}

Before proceeding with the calculation,
a comment is warranted on the choice of Killing vector from which to calculate the conserved charge.
To identify $\alpha$, one might try to again posit that $\partial_t$
is the appropriate Killing vector with which to calculate the mass
when it coincides with the time coordinate of the Rindler wedge.
In fact, a slight modification of this procedure is required:
one should scale the time coordinate by a factor of $m$ in order to
reproduce the zero-mass of AdS$_3$ as the black hole
horizon size is taken to zero \cite{Banados:1992wn}.
The mapping between the Class II geometry
and the Rindler geometry is
\begin{equation}
  \left(-1+\frac{R^2}{\ell^2}\right) =
  \frac{F(r)}{m^2\alpha^2\Omega(r,\psi)^2}
  \,,
  \qquad
  R\sinh\vartheta =
  \frac{r\sinh{\left(m\psi\right)}}{m\Omega(r,\psi)}
  \,, 
\label{eq:IILocaltoGlobal}
\end{equation}
where we must also scale the time coordinate
\begin{equation}
  \tilde{t} = \frac{t_\text{Rindler}}{m}
  \label{eq:ClassIItoRindlerTime}
\end{equation}
and set
\begin{equation}
  \alpha = \sqrt{1+m^2\mathcal{A}^2\ell^2}
  \,.
  \label{eq:IIrightnormalisation}
\end{equation}
Note that, as $m$ is taken to zero, the metric \eqref{eq:ClassIIrightmetcal} becomes
\begin{equation}
  ds^2 =
  \frac{1}{(1+\mathcal{A}r)^2}
  \left(
    \frac{r^2}{\ell^2}d\tilde{t}^2
    - \frac{\ell^2}{r^2} dr^2
    - r^2 d\psi^2
  \right)
  \,.
  \label{eq:IIrightAlimit}
\end{equation}
The transformation
\begin{equation}
  R =\frac{r}{1+\mathcal{A}r}
\end{equation}
reveals the parameter $\mathcal{A}$ to then be a gauge artifact;
the geometry is simply Poincar\'{e} AdS$_3$:
\begin{equation}
  ds^2 =
  -\frac{\ell^2}{R^2}dR^2
  +\frac{R^2}{\ell^2}
  \left(d\tilde{t}^2 - \ell^2 d\psi^2\right)
  \,.
  \label{eq:poincareAdS3}
\end{equation}
This conclusion is solidified by noticing that the stress
\begin{equation}
  \sigma = - \frac{m\mathcal{A}}{4\pi } \sinh\left(m\pi\right)
  \label{eq:IIrighttensionm}
\end{equation}
inducing the domain wall also vanishes in this limit.
Poincar\'{e} AdS$_3$ sets the zero-point energy of three-dimensional gravity
holographically \cite{Balasubramanian:1999re}.
As such, it is natural that the mass of the Class II solutions
should be calculated with respect to $\tilde{t}$, using equation \eqref{eq:ClassIItoRindlerTime}.
Though rather ad hoc, this prescription to calculate the mass with respect to $\tilde{t} = m^{-1}t_\text{Rindler}$
is the one usually followed for the BTZ black hole.
Indeed it has led to a number of interesting results including providing a first law \cite{Frassino:2015oca}.
We proceed to calculate the conserved mass with respect to $\partial_{\tilde{t}}$,
and leave a detailed justification for future work.

We expand the Class II metric given in table \ref{tab:3} using \eqref{eq:expan},
and seek Fefferman-Graham form order by order.
Again, a sensible redefinition of conformal factor
\begin{equation}
  F_1(\xi) = \frac{\Upsilon^{3/2}}{A\ell\omega(\xi)}
  \,,
  \label{eq:IIconformalfactor}
\end{equation}
where now
\begin{equation}
  \Upsilon = 1-A^2\ell^2(\xi^2-1)
  \,,
\end{equation}
simplifies proceedings.
The boundary metric is then
\begin{equation}
  \g0 =
  \frac{\omega(\xi)^2}{A^2}
  \left[
    \frac{dt^2}{\alpha^2}
    - \frac{A^2\ell^2  d\xi^2}{\left(\xi^2-1\right)\Upsilon^2}   \right]
  \,.
  \label{eq:IIg0}
\end{equation}
The leading correction in the Fefferman-Graham expansion is
\begin{multline}
  \g2 =
  -\frac{1}{2A^2\ell^2\alpha^2}
  \left[
    1+A^2\ell^2 
    - \left(\xi^2-1\right)\Upsilon^2
    \left(\frac{\omega'(\xi)}{\omega(\xi)}\right)^2
  \right] dt^2
  \\
  +
  \Bigg(
    \frac{1}{\left(\xi^2-1\right)\Upsilon}
    \left[
      \frac{1}{2\Upsilon}\left(1+A^2\ell^2\right)
      +\xi\left(1-3A^2\ell^2(\xi^2-1)\right)
      \frac{\omega'(\xi)}{\omega(\xi)}
    \right]
  \\
    +\frac{3}{2}\left(\frac{\omega'(\xi)}{\omega(\xi)}\right)^2
    -\frac{\omega''(\xi)}{\omega(\xi)}
  \Bigg) d\xi^2
  \,.
  \label{eq:IIg2}
\end{multline}
The non-zero components of the stress tensor are
\begin{multline}
  8\pi \omega(\xi)^2 \ell
  \langle T^{\tau}_{\tau} \rangle = 
    \frac{1}{2}\left(1+A^2\ell^2\right)
    + \xi\left(1-3A^2\ell^2(\xi^2-1)\right)\Upsilon
    \frac{\omega'(\xi)}{\omega(\xi)}
    \\
    +\left(\xi-1\right)\Upsilon^2
    \left[ 
      \frac{\omega''(\xi)}{\omega(\xi)}
      -
      \frac{3}{2}\left(\frac{\omega'(\xi)}{\omega(\xi)}\right)^2
    \right]
  \label{eq:IIstresstautau}
\end{multline}
and
\begin{equation}
  16\pi \omega(\xi)^2 \ell
  \langle T^{\xi}_{\xi} \rangle =
  -\left(1+A^2\ell^2\right)
  +\left(\xi-1\right)\Upsilon^2
  \left(\frac{\omega'(\xi)}{\omega(\xi)}\right)^2
  \,.
  \label{eq:IIstressxixi}
\end{equation}
The Ricci scalar of the boundary metric is
\begin{multline}
  R(\g0) = 
    \frac{2\Upsilon}{\ell^2\omega(\xi)^2}
    \Big(
      \xi \left(1-3A^2\ell^2(\xi^2-1)\right)
      \frac{\omega'(\xi)}{\omega(\xi)}
      \\
      +
      \left(\xi^2-1\right)\Upsilon
      \left[
        \frac{\omega''(\xi)}{\omega(\xi)}
        -\left(\frac{\omega'(\xi)}{\omega(\xi)}\right)^2
      \right]
    \Big)
    \,,
    \label{eq:IIriccig0}
\end{multline}
which satisfies the conformal anomaly relation \eqref{eq:anomaly}
with central charge $c\textsubscript{Virasoro}=3\ell/2$.

\begin{figure}[tb!]
  \centering
  \includegraphics[width=\textwidth]{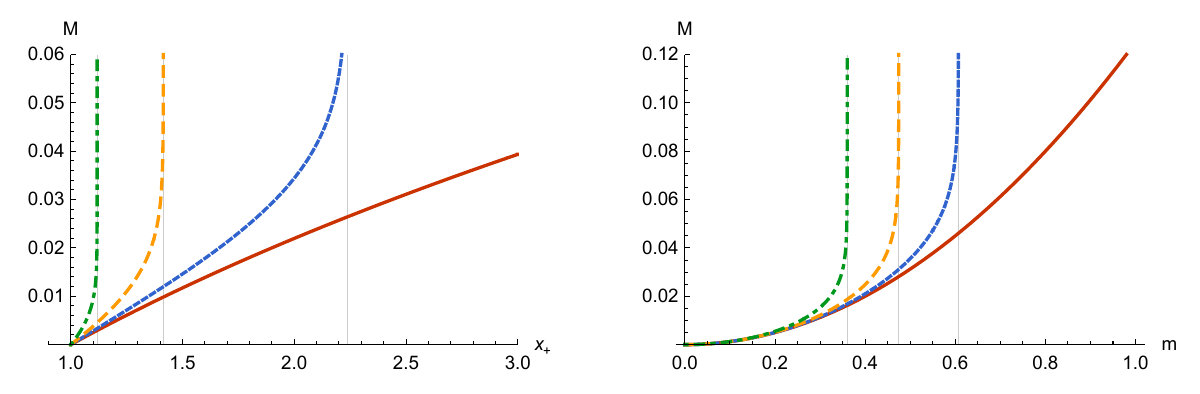}
  \caption[
    The holographic mass of the BTZ black-hole pushed by a strut,
    in its slow phase
  ]{
    The holographic mass $M$ of the BTZ black-hole
    pushed by a strut
    at various values of the acceleration parameter:
    $\mathcal{A}=0$ (solid red),
    $\mathcal{A}=0.5\ell^{-1}$ (dotted blue),
    $\mathcal{A}=\ell^{-1}$ (dashed orange),
    $\mathcal{A}=2\ell^{-1}$ (dot-dashed green).
  }
  \label{fig:IIrightmassplots}
\end{figure}

Accounting for the two patches needed
to cover the full region exterior to the horizon,
the mass associated with $\partial_{\tilde{t}}$ is given by
\begin{equation}
  M =
  2mA\int_{1}^{x_+}
  \sqrt{-\g0}\langle T^{\tau}_{\tau} \rangle d\xi
  \,,
\end{equation}
%
which when evaluated gives
\begin{equation}
  M =
  \frac{m}{8\pi \alpha}
  \sqrt{1+m^2\mathcal{A}^2\ell^2}
  \arctanh\left[
    \sqrt{1+m^2\mathcal{A}^2\ell^2}
    \tanh\left(m\pi\right)
  \right]
  \,.
  \label{eq:IIrightMassm}
\end{equation}
This is plotted at various values of $\mathcal{A}$ in figure 
\ref{fig:IIrightmassplots}, both as a function of $m$ and of $x_+$.
In the limit of vaishing $\mathcal{A}$
--- where the geometry reduces to that of the static BTZ black hole ---
the formula for mass reproduces the expected $M = m^2/8$.

\subsection{A BTZ black hole pulled by a wall (Class II\textsubscript{left})}
\label{sec:IIstring}

It is also possible to construct another one parameter extension
to the family of static BTZ black holes,
where the defect emerging from the horizon has a positive energy density.
As such, this solution is arguably more physical
than the solution of section \ref{sec:IIstrut}.

\begin{figure}[tb!]
  \centering
  \includegraphics[width=\textwidth]{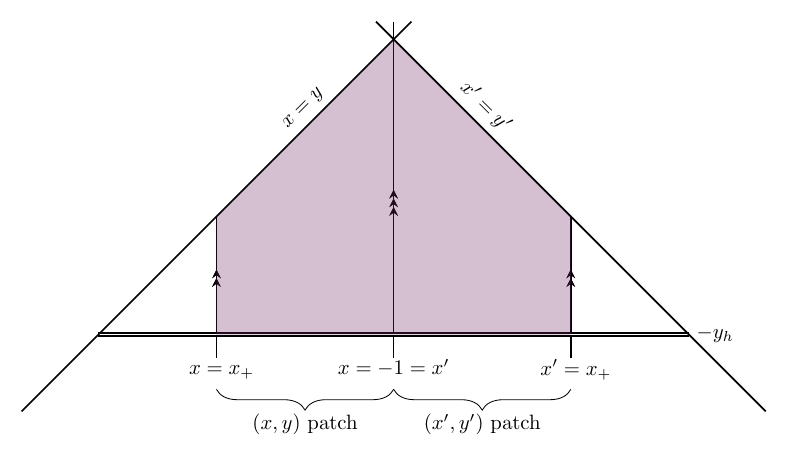}
  \caption[
    The two patches of spacetime
    used to construct the Class II\textsubscript{left} black hole
  ]
  {The two patches of type II spacetime
    used to construct a black hole
    with a pulling wall.}
  \label{fig:IIleftGlued}
\end{figure}

Starting from a patch of Class II spacetime \eqref{eq:metricxy2} with $x<-1$, 
as presented in table \ref{tab:3},
choose some value $x_+\in(-y_h,-1)$,
where $y_h=\sqrt{1+A^{-2}\ell^{-2}}$,
and define a patch with $x\in(x_+,-1]$.
Glue two copies of this patch, mirroring along both $x=x_+$ and $x=-1$.
The identifications are shown in figure \ref{fig:IIleftGlued}.
The $x=-1$ axis of the newly formed spacetime is regular.
Along $x=x_+$, one finds a domain wall
of positive tension $\sigma=A(4\pi )^{-1}\sqrt{Q(x_+)}$.
In contrast to the soluton with a strut,
here there is always one---and only one---compact horizon present 
at $y=-y_h$, 
regardless of the values of $A$ and $x_+$.
As $|x_+|\rightarrow y_h$,
instead of the system forming a non-compact horizon.
the black hole horizon merges with the conformal boundary,

To cast the metric in more familiar coordinates,
take $x=-\cosh(\psi/K)$, 
where $K=\pi/\arcosh(-x_+)$.
Defining dimensionful coordinates $\rho=-(Ay)^{-1}$ and $t=\alpha A^{-1}\tau$,
the metric becomes
\begin{align}
\begin{split}
  ds^2 &= 
  \frac{1}{\left[ 1 - A\rho\cosh\left(\psi/K\right) \right]^2}
  \left[
    f(\rho) \frac{dt^2}{\alpha^2}
    - \frac{d\rho^2}{f(\rho)}
    - \rho^2 \frac{d\psi^2}{K^2}
  \right]
  \,,
  \\
  f(\rho) &= -1 + (1+A^2\ell^2) \rho^2/\ell^2
  \,.
\end{split}
  \label{eq:IIleftmetricrhosig}
\end{align}
Yet again, we anticipate a rescaling of $t$ by $\alpha$.
The range of $\psi$ is $(-\pi,\pi)$.
This $(\rho,\psi)$ patch covers both $(x,y)$ coordinate patches;
it covers the entire region region between the black hole horizon
and the conformal boundary
(the region shaded purple in figure \ref{fig:IIleftGlued}).
The coordinate $\rho$ is everywhere positive in this domain.
As long as the acceleration parameter $A$ is sufficiently small,
it is possible to set the magnitude of $x_+$ as large as one wishes.
We thus have two conditions on $K$ for this solution:
$K>0$ and $\pi/K<\arcosh(y_h)$.

The semiaxis $\psi=0$ is regular,
while the domain wall,
which now lies along $\psi=\pm\pi$,
has tension
\begin{equation}
  \sigma = \frac{A}{4\pi } \sinh\left(\frac{\pi}{K}\right)\,.
\end{equation}

Much like the solutions with a strut considered in section \ref{sec:IIstrut}, 
the solutions we have constructed in this section
comprise a one parameter extension to the family of static BTZ black holes.
One may demonstrate this in a similar way.
Take the metric \eqref{eq:IIleftmetricrhosig}
and again make the parameter redefinitions
$K=m^{-1}$ and $A=m\mathcal{A}$.
Also, make the coordinate rescalings $r=m\rho$ and $\tilde{t}=m^{-1}t$.
The metric becomes
\begin{align}
\begin{split}
  ds^2 &= 
  \frac{1}{\Omega(r,\psi)^2}
  \left[
    F(r) \frac{d\tilde{t}^2}{\alpha^2}
    - \frac{dr^2}{F(r)}
    - r^2 d\psi^2
  \right]
  \,,
  \\
  F(r) &= -m^2(1-\mathcal{A}^2r^2) + \frac{r^2}{\ell^2}
  \,,
  \\
  \Omega(r,\psi) &= 1-\mathcal{A}r\cosh(m\psi)
  \,.
\end{split}
\label{eq:metricIIstringm}
\end{align}
We now have the conditions
$0\leq m\sinh\left(m\pi\right)<\left(\mathcal{A}\ell\right)^{-1}$.
Note that, much as for the solution with a strut,
the point $m=0$ in parameter space was inaccessible
using the parameterisation \eqref{eq:IIleftmetricrhosig}.
The black hole horizon lies at $r=m\ell(1+m^2\mathcal{A}^2\ell^2)^{-1/2}$.
The tension in the domain wall is an increasing function of $m$:
\begin{equation}
  \sigma = \frac{m \mathcal{A}}{4\pi } \sinh\left(m\pi\right)
  \,.
  \label{eq:IIlefttensionm}
\end{equation}
It is interesting to note that the bounds on $m$ translate
into bounds on the induced defect stress:
\begin{equation}
    0 \leq \sigma < \frac{1}{4\pi \ell}
    \,.
\end{equation}

To understand the geometry, we map onto a subset of global AdS$_3$
and plot the geometry by compactifying the spatial section.
This process is described in appendix \ref{app:map},
with the result being figure \ref{fig:3DIIstring}.
One finds a spacetime qualitatively similar to the BTZ black hole,
(cf. figure \ref{fig:3Dbtz}), but with non-zero tensile force present 
in the identification surface.
The bifurcate Killing horizon generated by $\partial_t$
has a compact bifurcation surface, with the topology of a circle.
\begin{figure}[tb!]
  \centering
  \includegraphics[width=0.45\textwidth]{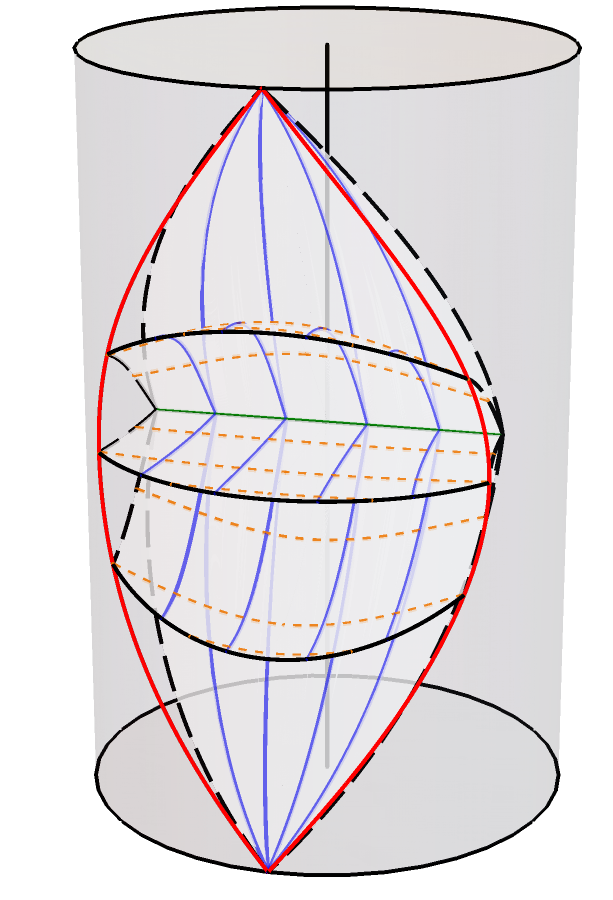}
  \hfill
  \includegraphics[width=0.45\textwidth]{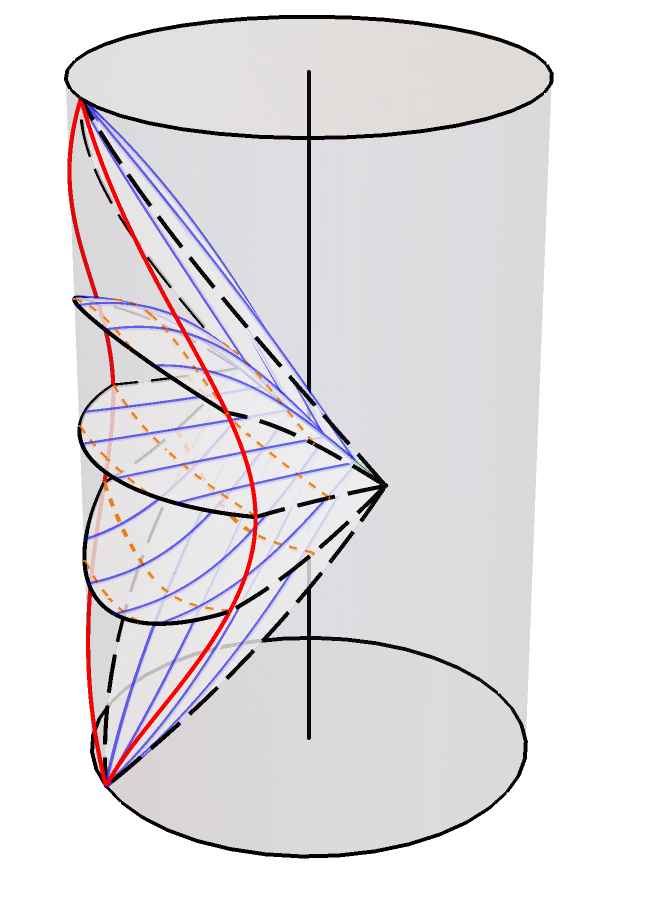}
  \caption[The Class II\textsubscript{left} black hole
    ]{
      The Class II\textsubscript{left} black hole.
      Several surfaces of constant $t$ are shown;
      those of large positive and negative $t$ outline the event horizon.
      The surface $x=x_+$ -- shown in dashed black -- is identified
      with its partner within each time-slice.
      The bifurcation surface of the Killing horizon is shown in green
      and has topology $S^1$.
      Lines of constant $x$ and $y$ are shown in solid blue and dashed orange
      respectively.
      The classically accessible region of the conformal boundary is
      delimited in red.
    }
  \label{fig:3DIIstring}
\end{figure}

\subsubsection{The holographic mass}
As the geometry never forms an acceleration horizon,
one is free to calculate the holographic mass over the entire range of $K$
(equivalently $m$).
We again make the decision to calculate the conserved charge with respect to
$\partial_{\tilde{t}}$, where $\tilde{t} = m^{-1}t_{\text{Rindler}}$.
This is justified similarly to the argument given in 
subsection \ref{subsec:IIstrutMass}.
The transformations as written in \eqref{eq:IILocaltoGlobal}
hold in the present case
(though of course one must use the appropriate $\Omega$),
leading to the same value of $\alpha$ \eqref{eq:IIrightnormalisation}.
The procedure to construct the holographic stress tensor
is then identical to the one performed in section \ref{sec:IIstrut};
we will not repeat the details.
In particular, the expressions for
the Fefferman-Graham expansion
\eqref{eq:IIconformalfactor},
\eqref{eq:IIg0},
\eqref{eq:IIg2};
the stress tensor
\eqref{eq:IIstresstautau},
\eqref{eq:IIstressxixi};
and
the Ricci scalar of the boundary metric
\eqref{eq:IIriccig0}
are all identical.

\begin{figure}[tb!]
  \centering
  \includegraphics[width=\textwidth]{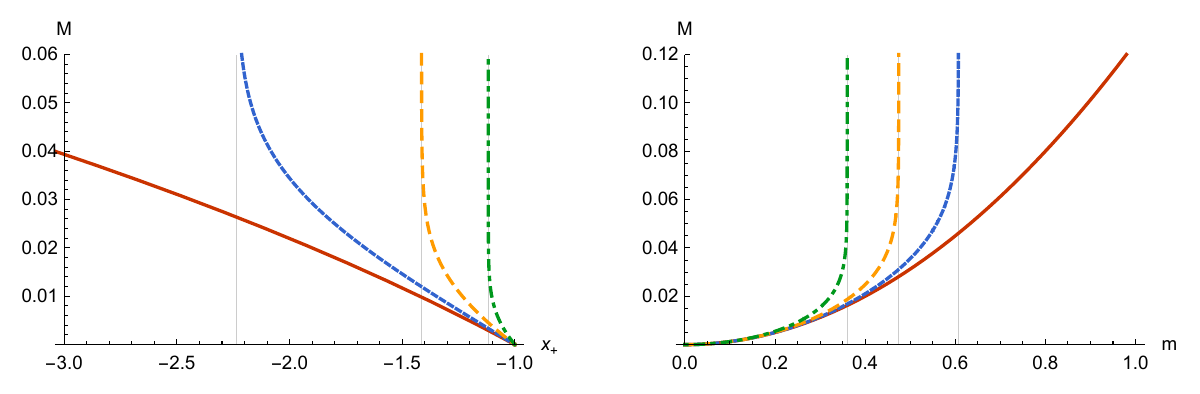}
  \caption[
    The holographic mass of the BTZ black-hole pulled by a wall
  ]{
    The holographic mass $M$
    of the BTZ black-hole pulled by a wall
    at various acceleration parameters:
    $\mathcal{A}=0$ (solid red),
    $\mathcal{A}=0.5\ell^{-1}$ (dotted blue),
    $\mathcal{A}=\ell^{-1}$ (dashed orange),
    $\mathcal{A}=2\ell^{-1}$ (dot-dashed green).
  }
  \label{fig:IIleftmassplots}
\end{figure}

The limits for the mass integration must be updated for the present case,
\begin{equation}
  M =
  2A\int_{x_+}^{-1}
  \sqrt{-\g0}\langle T^{\tau}_{\tau} \rangle d\xi
  \,,
\end{equation}
but the evaluated value is identical to the mass of the black hole 
with a strut:
\begin{equation}
  M =
  \frac{m}{8\pi \alpha}
  \sqrt{1+m^2\mathcal{A}^2\ell^2}
  \arctanh\left[
    \sqrt{1+m^2\mathcal{A}^2\ell^2}
    \tanh\left(m\pi\right)
  \right]
  \,.
  \label{eq:IIleftmassm}
\end{equation}
The mass is plotted at various values of $\mathcal{A}$ in figure 
\ref{fig:IIleftmassplots}, both as a function of $m$ and of $x_+$.
As $\mathcal{A}\rightarrow 0$,
$M$ approaches the mass of a standard BTZ black hole
\cite{Balasubramanian:1999re} $M\textsubscript{BTZ} = m^2/8$
appropriately.
As $m\rightarrow 0$, the expression \eqref{eq:IIleftmassm} for $M$ vanishes.
In this limit, simple coordinate transformation
\begin{equation}
  R =\frac{r}{1-\mathcal{A}r}
\end{equation}
applied to the line element \eqref{eq:metricIIstringm}
recovers \eqref{eq:poincareAdS3},
the Poincar\'{e} patch of AdS$_3$
with a periodically identified spatial coordinate.
In consonance with \eqref{eq:IIleftmassm},
this geometry has vanishing mass \cite{Balasubramanian:1999re}.

\section{A BTZ black hole pulled by a wall (Class I\textsubscript{C})}
\label{sec:IC}

We now return to the Class I solutions to explore a novel BTZ accelerating
black hole that has only a limited allowed parameter range and is disconnected 
from the non-accelerating BTZ black hole.

Consider a Class I geometry in the rapid (or saturated) phase, so that 
$A^2\ell^2\geq 1$. For positive $y$, there is a Killing horizon at 
$y_h=\sqrt{1-A^{-2}\ell^{-2}}$. Now choose $x_+\in(y_h,1)$, so that
the $y$ coordinate lies in the range $(y_h,x)$ for $x\in(x_+,1)$. As usual, we 
glue two copies of this chart along both $x=x_+$ and $x=1$. Diagrammatically, 
we are joining two copies of the most heavily shaded region in figure 
\ref{fig:I}(Left) where the relevant value of $x_+$ is denoted by $C$.
There is no tension along the gluing surface $x=1$ as usual, and
along $x=x_+$, there is a domain wall with positive tension
\begin{equation}
\sigma =\frac{A}{4\pi }\sqrt{1-x_+^2} \,.
\end{equation}
This positive tension domain wall extends out from the (now) 
compact horizon to the conformal boundary at $y=x$.

As before, we transform to the more intuitive coordinate system
\begin{equation}
\tau=\frac{At}{\alpha} \,,\qquad
y=\frac{1}{A\rho} \,,\qquad
x=\cos\left(\phi/K\right)   \,,
\end{equation}
where $K=\pi/\arccos\left(x_+\right)$. Since for these solutions $x_+\in(y_h,1)$ 
and $y_h\geq0$, we have that $K>\pi/\arccos\left(y_h\right)>2$. $K$ can only 
approach its minimal value of $2$ as the solution becomes close to saturated.
The metric takes the form
\begin{equation}
\begin{split}
ds^2 &=
\frac{1}{\left[ A\rho\cos\left(\frac{\phi}{K}\right) - 1 \right]^2}
\left( f(\rho)\frac{dt^2}{\alpha^2} -\frac{d\rho^2}{f(\rho)}
-\rho^2\frac{d\phi^2}{K^2} \right) \,,\\
f(\rho) &= 1-(A^2\ell^2-1) \rho^2/\ell^2 \,.
\end{split}
\label{eq:ICmetricrho}
\end{equation}
The range of $\phi$ is $(-\pi,\pi)$, covering both $(x,y)$ patches.
The horizon lies at $\rho_h=(Ay_h)^{-1}$, with the conformal boundary 
$\rho_\text{conf.}=(A\cos(\phi/K))^{-1}$ satisfying 
$0<\rho_\text{conf.}<\rho<\rho_h$. With this parameterisation, the domain 
wall lies on the line $\phi=\pm\pi$ and has tension
\begin{equation} 
\sigma = \frac{A}{4\pi }\sin\left(\frac{\pi}{K}\right) \,.
\end{equation}
The tension is bounded above by
\begin{equation}
\sigma_\text{max} = \frac{A}{4\pi } \,,
\end{equation}
which is only approachable for saturated or (nearly saturated) solutions 
with values of $K$ very close to $2$.
It is monotonically decreasing with $K$ and bounded below by zero.

\begin{figure}[t!]
\centering
\includegraphics[width=0.5\textwidth]{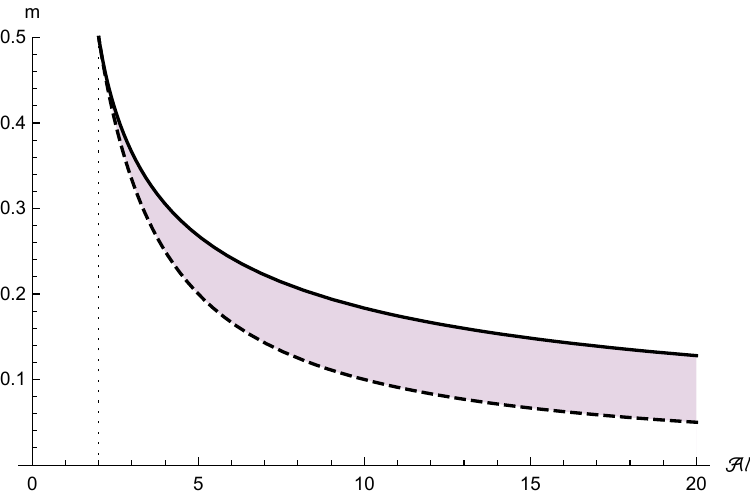}
\caption{The parameter space of the I\textsubscript{rapid,C} solution (shaded purple).}
\label{fig:IC_mrange}
\end{figure}

An alternative parameterisation of the solution is available via the substitutions
\begin{equation}
m=\frac{1}{K}  \,,\qquad \mathcal{A} = \frac{A}{m} \,; \qquad
\tilde{t} = \frac{t}{m} \,, \qquad r=m\rho \,.
\end{equation}
The metric becomes
\begin{equation}
\begin{split}
  ds^2 &=
    \frac{1}{\Omega(r,\phi)^2}
    \left(
      F(r)\frac{d\tilde{t}^2}{\alpha^2}
      -\frac{dr^2}{F(r)}
      -r^2 d\phi^2
    \right)
    \,,\\
  F(r) &=
    -m^2(\mathcal{A}^2r^2-1) + \frac{r^2}{\ell^2}
    \,,\\
  \Omega(r,\phi) &=
    \mathcal{A}r\cos\left(m\phi\right) - 1
    \,. 
\end{split}
\end{equation}
The bounds on $K$ translate into bounds on $m$:
\begin{equation}
  0<m<\frac{1}{\pi}\arccos\left(y_h\right)<\frac{1}{2}
  \,.
  \label{eq:IC_mbounds}
\end{equation}
Since $m$ is bounded above by one half,
no geometries of this type are possible with $\mathcal{A}\ell<2$.
Thus, this corner of solution space, while representing an 
accelerating BTZ black hole, is not connected to the canonical
BTZ solution as $A\to0$ is not allowed for these solutions.
In fact, after the reparameterization, $y_h$ now depends on $m$.
This makes the second inequality in \eqref{eq:IC_mbounds} slightly tricky;
it is more easily expressed as a bound on $\mathcal{A}\ell$:
\begin{equation}
  \frac{1}{m} \leq \mathcal{A}\ell < \frac{1}{m\sin\left(m\pi\right)}
  \,.
  \label{eq:mscAbounds}
\end{equation}
This effective upper bound on $m$ combines with
the lower bound provided by the rapid acceleration condition
to form an allowed range of $m$ values for a given $\mathcal{A}\ell$,
shown in figure \ref{fig:IC_mrange}.

The conformal boundary lies at
$r_\text{conf.}=(\mathcal{A}\cos(m\phi))^{-1}\geq\mathcal{A}^{-1}$.
The minimal value of $\mathcal{A}^{-1}$
is achieved only at a single point:
the point on the boundary for which $\phi=0$.
For a rapid ($m\mathcal{A}\ell>1$) solution, the horizon lies at
\begin{equation}
r_h = \frac{m\ell}{\sqrt{m^2\mathcal{A}^2\ell^2-1}}
\end{equation}
and satisfies $r_h > \mathcal{A}^{-1}$.
As the solution becomes saturated ($m\mathcal{A}\ell=1$), the horizon 
moves out to infinity and these coordinates become less useful.

\begin{figure}[t!]
  \centering
  \includegraphics[width=0.5\textwidth]{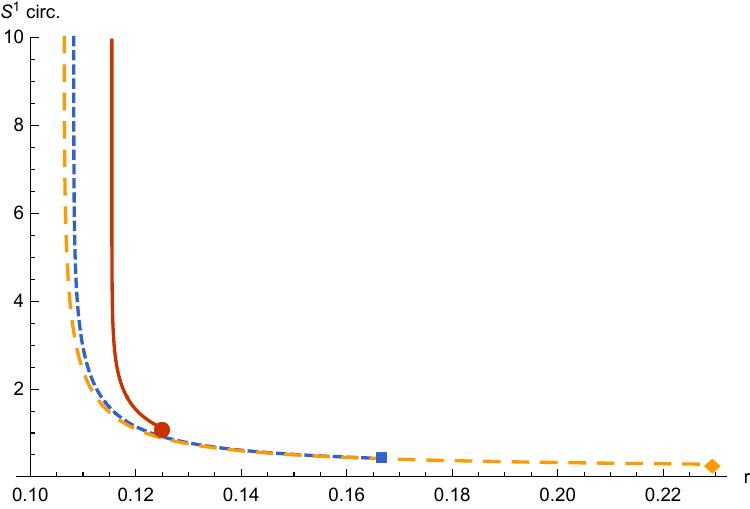}
  \caption[
    The ``circumference'' of loops of constant $r$ in the I\textsubscript{rapid,C} solution
  ]{
    The ``circumference'' of loops of constant $r$.
    $\ell$ is set to unity with $\mathcal{A}=10$.
    Three values of $m$ ($6^{-1}$, $8^{-1}$, and $9^{-1}$) are shown
    (in solid red, dotted blue, and dashed orange respectively).
    The value at the horizon radius $r=r_h$ is marked by a shape.
  }
  \label{fig:IC_rings}
\end{figure}

One might be concerned that $r$ is decreasing
as one nears the conformal boundary.
The length of a closed ring of constant $r$ is
\begin{equation}
  \int_{-\pi}^{\pi}
  \frac{r}{\mathcal{A}r\cos\left(m\phi\right)-1}
  d\phi
  =
  \frac{r}{m\sqrt{\mathcal{A}^2r^2-1}}
  \arctanh
  \left[
    \sqrt{\frac{\mathcal{A}r+1}{\mathcal{A}r-1}}
    \tan\left(\frac{m\pi}{2}\right)
  \right]
  \,.
  \label{eq:IC_ringcircumference}
\end{equation}
This function is monotonically decreasing with $r$,
as shown in figure \ref{fig:IC_rings}.
A ring close to the horizon
is smaller than a ring close to the boundary;
the system is behaving in an intuitive fashion,
despite the unintuitive metric.

To better understand the solution,
we can map it to a subset of global AdS$_3$
and plot it by compactifying the spatial section.
This process is described in appendix \ref{app:map},
with the result being figure \ref{fig:3DIC}.
One finds a result qualitatively similar to the BTZ black hole,
(cf. figure \ref{fig:3Dbtz}), but with non-zero stress present 
in the identification surface.
This is the key to understanding the I\textsubscript{rapid,C}
and I\textsubscript{saturated,E,$y>0$} solutions.
The static BTZ black hole is constructed by taking the AdS$_3$ Rindler wedge
\eqref{eq:RindlerWedge}
and identifying complete orbits of the Killing vector generating rotations,
(see appendix \ref{app:staticBTZ} for a short review).
The Class I\textsubscript{C} black hole is constructed similarly,
though one is identifying instead surfaces of constant $x$.
\begin{figure}[t!]
  \centering
  \includegraphics[width=0.45\textwidth]{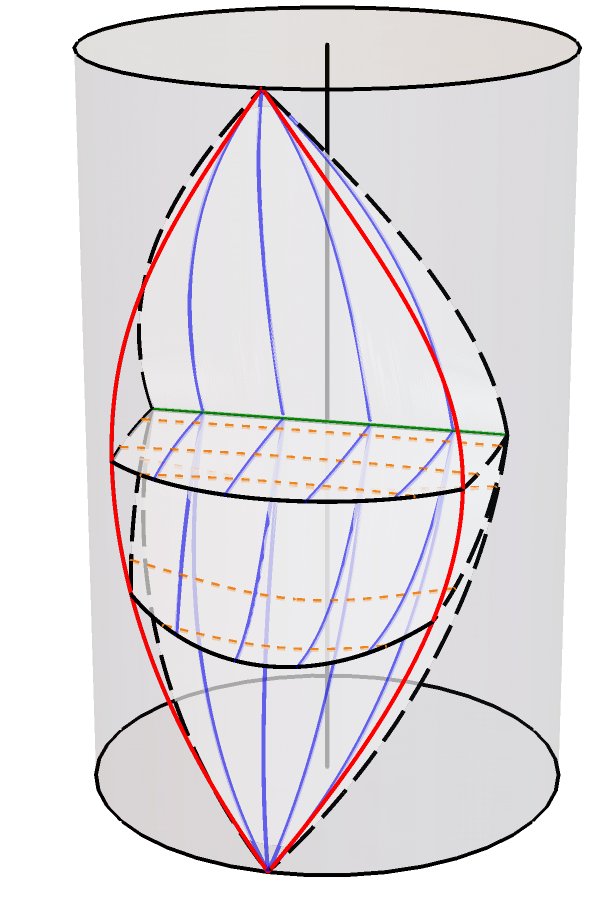}
  \hfill
  \includegraphics[width=0.45\textwidth]{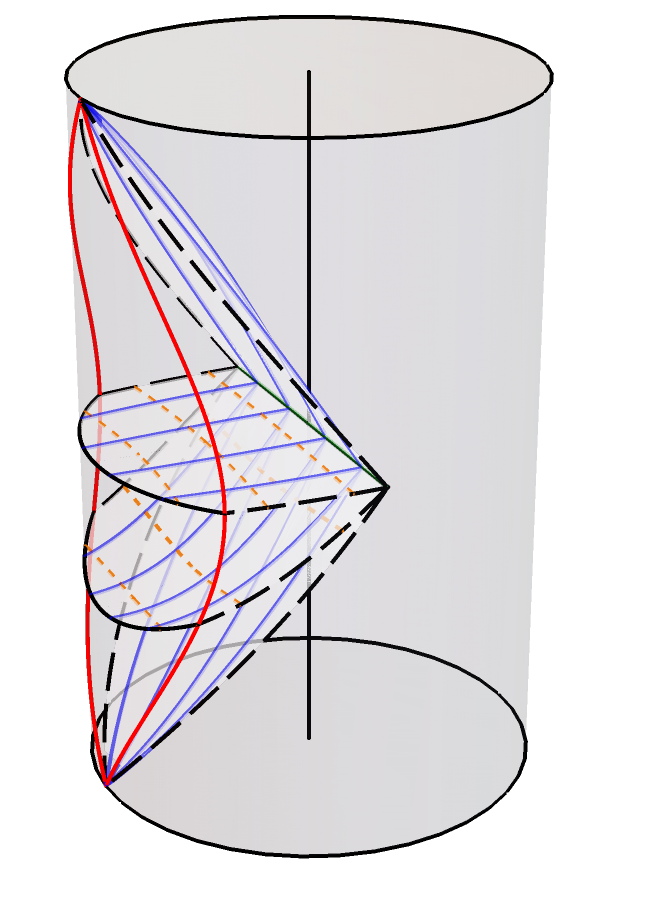}
  \caption[The Class I\textsubscript{rapid,C} black hole
    ]{
      The Class I\textsubscript{rapid,C} solution, as a portion of global AdS$_3$.
      Several surfaces of constant $t$ are shown.
      The horizon is shown by the surfaces at early and late times,
      with the bifuration surface shown in green.
      The lines $x=x_+$ and $x'=x_+$ within each time-slice
      (shown in long-dashed black) are identified,
      which imbues the birfurcation surface with the topology of a circle.
      Lines of constant $x$ are shown in blue,
      with lines of constant $y$ shown in dashed orange.
      The classically accessible subset of the global boundary 
      is delimited in red.
      To guide the eye, the locus of the cylinder is shown in solid black.
    }
  \label{fig:3DIC}
\end{figure}

\subsection{Physical properties}
\label{sec:IC_properties}
A short discussion of the time coordinate we have written down for the solution
is warranted.
The coordinate $t$ of \eqref{eq:ICmetricrho} is in fact the Rindler time
of the gauge \eqref{eq:RindlerWedge}.
The explicit mapping of the Class I\textsubscript{C} geometry
to the planar BTZ geometry \eqref{eq:RindlerWedge} is given by
\begin{equation}
  \left(-1+\frac{R^2}{\ell^2}\right) =
  \frac{F(r)}{m^2\alpha^2\Omega(r,\phi)^2}
  \,,
  \qquad
  R\sinh\vartheta =
  \frac{r\sin{\left(m\phi\right)}}{m\Omega(r,\phi)}
  \,,
\label{eq:ICLocaltoRindler}
\end{equation}
where the time coordinates are related by
\begin{equation}
  \tilde{t} = \frac{t_{\text{Rindler}}}{m}
  \,.
\end{equation}
The above transformation necessitates
\begin{equation}
    \alpha=\sqrt{m^2\mathcal{A}^2\ell^2-1}
    \,.
    \label{eq:ICnormalisation}
\end{equation}
There are then two obvious candidates,
$\partial_t$ and $\partial_{\tilde{t}}$,
for the Killing vector 
from which to compute physical quantities such as the mass and temperature.
It is not clear which, if either, of these is the correct choice,
as the solution is not smoothly connected to any familiar ones.
In this subsection we compute quantities with respect to
the Rindler time Killing vector $\partial_t$.

\begin{figure}[t!]
  \centering
  \includegraphics[width=0.5\textwidth]{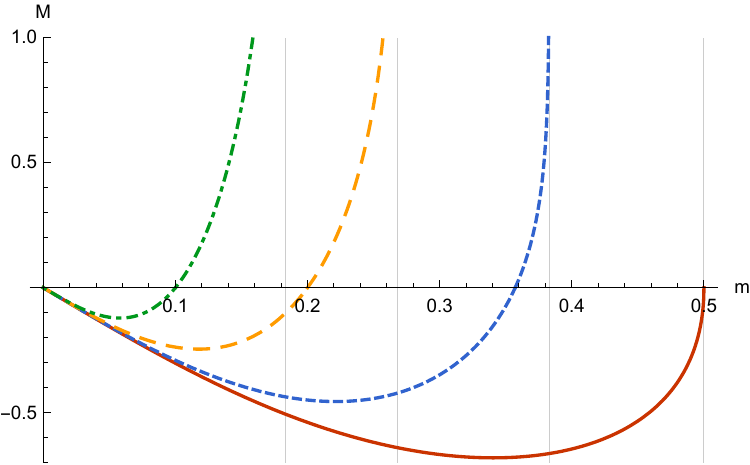}
  \caption[
    The holographic mass of the I\textsubscript{rapid,C} solution
  ]{
    A plot of $8\pi \alpha M$ at various values of $\mathcal{A}\ell$:
    $\mathcal{A}\ell=2.0$ (solid red),
    $\mathcal{A}\ell=2.8$ (dotted blue),
    $\mathcal{A}\ell=5.0$ (dashed orange),
    $\mathcal{A}\ell=10$ (dot-dashed green).
    The grey vertical lines denote the asymptotes where the divergences
    $\mathcal{A}\ell = (m\sin(m\pi))^{-1}$ are met.
    Negative values of $M$ are forbidden by the lower bound
    $m\geq(\mathcal{A}\ell)^{-1}$. 
  }
  \label{fig:IC_massm}
\end{figure}

Since the geometry possesses one isolated, compact horizon,
the calculation of holographic mass goes through similarly
to those for the accelerating particle solutions constructed
in sections \ref{sec:Istring} and \ref{sec:Istrut}.
The limits of integration need updating
for the I\textsubscript{rapid,C} solution:
\begin{equation}
  M =
  2m\mathcal{A}\int_{x_+}^{1}
  \sqrt{-\det\g0}
  \left\langle T_\tau^\tau\right\rangle
  d\xi
  \,,
  \label{eq:ICmassIntegral}
\end{equation}
where recall that now $x_+>0$ and $m\mathcal{A}\ell\geq1$.
The integral may be evaluated to obtain
\begin{equation}
  M =
  \frac{1}{8\pi \alpha}
  \sqrt{m^2\mathcal{A}^2\ell^2-1}
  \arcoth
  \left[
    \frac{\cot\left(m\pi\right)}{\sqrt{m^2\mathcal{A}^2\ell^2-1}}
  \right]
  \,.
\end{equation}
As we approach saturation, $m\mathcal{A}\ell\rightarrow 1$ from above,
the mass vanishes.
The mass, with the normalisation \eqref{eq:ICnormalisation},
is plotted against $m$ in figure \ref{fig:IC_massm}.
Though it might appear that negative values of $M$ are possible for small $m$,
in fact, negative masses are forbidden by the condition that
the solution is rapid or saturated;
the roots of $M$ occur exactly at the saturation point
$m=(\mathcal{A}\ell)^{-1}$.
The mass spectrum of the solutions is non-negative,
although there is a narrow range of allowed $m$ for each $\mathcal{A}\ell$.
Within the allowed parameter space,
the holographic mass is monotonically increasing with $m$,
and can become unboundedly large.

\begin{figure}[t!]
  \centering
  \includegraphics[width=0.5\textwidth]{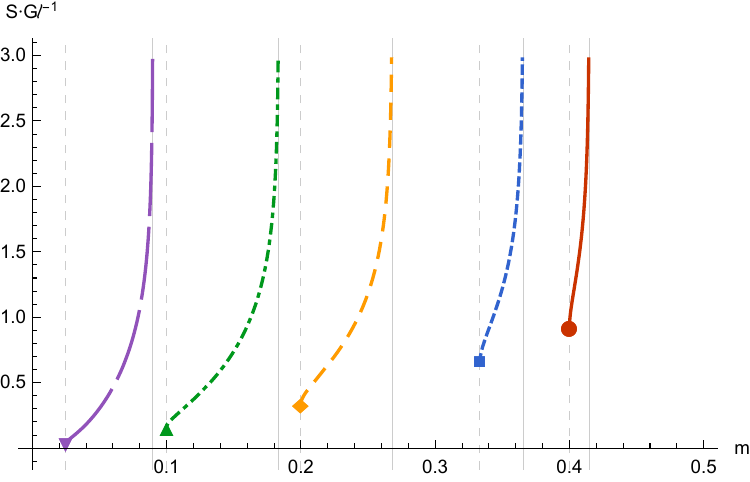}
  \caption[
    The entropy of the I\textsubscript{rapid,C} solution
  ]{
    The entropy $\ell^{-1}S$ of the I\textsubscript{rapid,C} solution
    at various values of $\mathcal{A}\ell$:
    $\mathcal{A}\ell=2.5$ (solid red),
    $\mathcal{A}\ell=3.0$ (dotted blue),
    $\mathcal{A}\ell=5.0$ (dashed orange),
    $\mathcal{A}\ell=10$ (dot-dashed green),
    $\mathcal{A}\ell=40$ (long-dashed purple).
    A solid grey vertical line denotes an asymptote where the divergence
    $\mathcal{A}\ell = (m\sin(m\pi))^{-1}$ is met.
    A dashed grey line dentotes the value of $m$ for a massless solution;
    the entropy at this point is marked with a shape.
  }
  \label{fig:IC_entropy}
\end{figure}

Calculating the minimal value of \eqref{eq:IC_ringcircumference},
the horizon is seen to have entropy
\begin{equation}
  S =
    \ell\arctanh
    \left[
      m\mathcal{A}\ell\left(1+y_h\right)
      \tan
      \left(
        \frac{m\pi}{2}
      \right)
    \right]
  \,.
\end{equation}
The entropy is monotonically increasing in $m$,
with the massless solution attaining the minimal entropy
\begin{equation}
  S_{M=0} =
    \ell\arctanh
    \left[  \tan\left(\frac{\pi}{2\mathcal{A}\ell}\right)  \right]
  \,.
\end{equation}
The entropy is plotted against $m$, for various values of $\mathcal{A}\ell$,
in figure \ref{fig:IC_entropy}.
In the figure, the entropy of the massless solution is marked by a shape.
The entropy is monotonically increasing with the mass
and diverges as $m$ approaches its supremum.

By regularity of the Euclidean section,
the horizon temperature is
\begin{equation}
  T
  = \frac{\left\vert f'(r_h) \right\vert}{4\pi\alpha}
  = \frac{1}{2\pi\ell\alpha}\sqrt{m^2\mathcal{A}^2\ell^2-1}
  \,.
\end{equation}
It is interesting to note that either of the possible normalisations
$\partial_t$ and $\partial_{\tilde{t}}$ yield
a temperature independent of the acceleration parameter $\mathcal{A}$.
At the saturation point $m\mathcal{A}\ell=1$, where the solution is massless,
the system approaches absolute zero.

\section{Class III solutions}
\label{sec:III}

\begin{figure}[t!]
  \centering
  \includegraphics[width=0.5\textwidth]{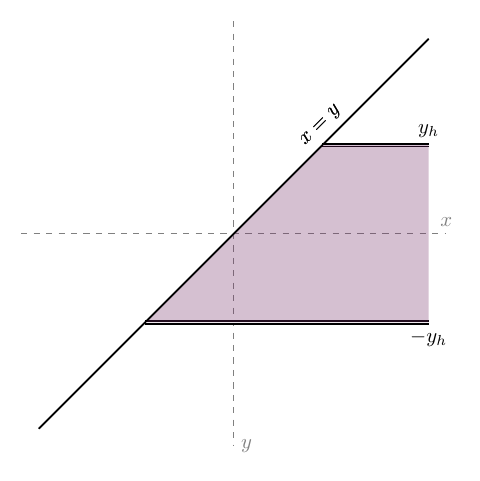}
  \caption[
  Coordinate ranges for the Class III solution
  ]{
  Coordinate ranges for the Class III solution.
  }
  \label{fig:IIIrapid}
\end{figure}

As stated in section \ref{sec:solutions},
if $A^2\ell^2\geq1$ then $P$ is everywhere non-positive.
We thus restrict to the case where $A^2\ell^2<1$.
There are then two disconected, non-compact horizons in the spacetime
at $y=\pm y_h$, where $y_h \equiv \sqrt{A^{-2}\ell^{-2}-1}$.
Solutions of Class III have no roots for $Q(x)$
and so the acceptable range of $x$ is $x>y>-y_h$.
The $(x,y)$ parameter space is shown in figure \ref{fig:IIIrapid}.
Since there are no roots of $Q$ at which to form a regular semi-axis,
it is not possible to form a single-wall solution
with the interpretation of $x$ as an angular coordinate.
While it is possible to form a black hole solution with two domain walls,
we do not do so here.

\begin{figure}[tb!]
  \centering
  \includegraphics[width=0.45\textwidth]{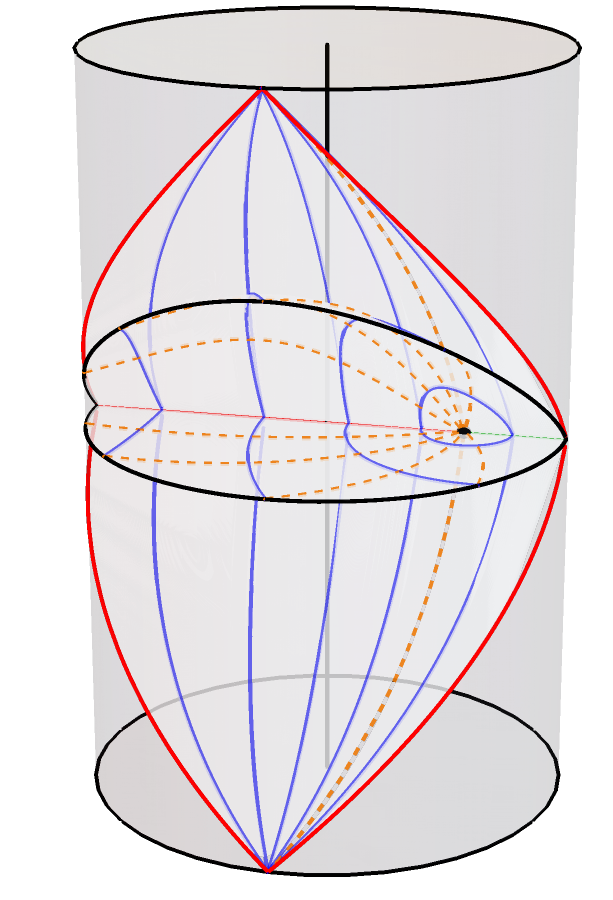}
  \hfill
  \includegraphics[width=0.45\textwidth]{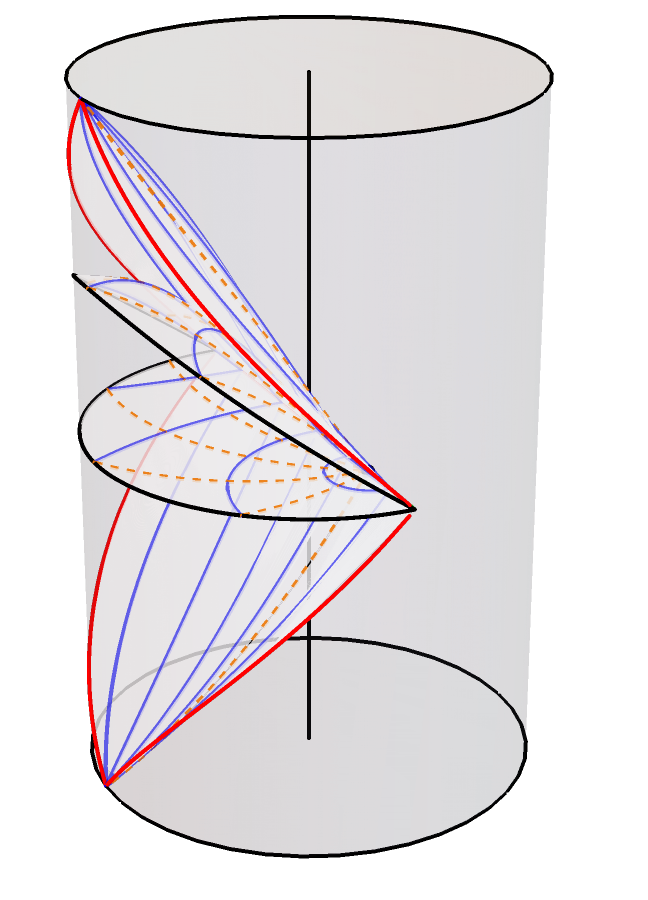}
  \caption{
    The Class III solution.
    Several surfaces of constant $\tau$ are shown,
    with those at very early and very late times outlining the two disjoint bifurcate Killing horizons.
    The red bifurcation surface is $y=-y_h$, while the green bifurcation surface is $y=+y_h$.
    The black dot denotes the point $x\rightarrow\infty$.
    Lines of constant finite $x$ are shown in solid blue, with lines of constant
    $y$ shown in dashed orange.
  }
  \label{fig:3DIII}
\end{figure}

Although it is not possible to make a single-wall solution with a periodic $x$ coordinate,
One could --- if one desired --- cut the patch denoted in figure \ref{fig:3DIII}
along some line $x=x_+$
which connects the $y=-y_h$ horizon to either
the conformal boundary (for $x_+<y_h$)
or
the $y=-y_h$ horizon (for $x_+>y_h$).
Identifying two copies for the remaining space along the cut still gives a solution,
although $x$ is not periodic.
Four such solutions are possible, depending on a choice of signs for $x_+-y_h$ and $x-x_+$.
This solution is a stark departure from the ones we have considered in previous sections.
In particular,
the resulting space is some subset of \textit{two} copies of the anti-de Sitter covering space,
$\text{AdS}_3\times\text{AdS}_3$;
the solution is no-longer guaranteed to be embeddable within a single copy of AdS$_3$.
This is a ``braneworld solution'',
with the brane at $x=x_+$ dividing the two copies of global space.
We will not comment on the detailed properties of these solutions here,
but it is worth acknowledging as a curiosity.

\section{Conclusions}
\label{sec:discussion}

In this paper, we have constructed within three-dimensional general relativity 
a broad family of solutions resembling the four-dimensional C-metric,
showing that the set of possible geometries is much richer than previously
acknowledged in the literature. We found solutions describing both point masses
(conical defects) and black holes, computing the mass of each holographically 
for phases without non-compact horizons.
It is interesting to note that even with the presence of domain walls,
a qualitative property of the theory's spectrum is unchanged:
the theory is ``gapped'', with point particle and black hole solutions
occupying the negative and positive-mass sectors, respectively.

In section \ref{sec:accptcle}, we found solutions representing a point particle
accelerated by a domain wall or strut. We demonstrated the existence of both 
rapidly and slowly accelerating phases, showing that there is a region of 
parameter space for which the formation of a non-compact acceleration 
horizon may be circumvented. Much of the behaviour of these solutions is
reminiscent of the four-dimensional accelerated black hole, described by the 
C-metric with negative cosmological constant. Both the three and 
four-dimensional solutions possess slowly and rapidly accelerating phases.
In their slowly accelerating phases, both solutions describe an isolated mass
held a fixed distance from the origin of pure anti-de Sitter space by a topological 
defect. This defect fixes the object to a worldline with a constant four-acceleration.
The rapidly accelerating phases of the four-dimensional C-metric and 
three-dimensional conical defect solutions also share similarities.
In particular -- at least for small black holes in four dimensions -- both 
geometries are approximately Rindler \cite{Dias:2002mi,Griffiths:2006tk}.

On the other hand, the three-dimensional point particles possess novel
phases which are not shared by their four-dimensional cousins.
Consider for example the Class I\textsubscript{rapid,A} single-wall solution, 
describing a rapidly accelerating particle pulled by a domain wall.
While for light particles the structure is similar to a small, accelerating black 
hole in four dimensions, as particle mass increases a phase transition occurs 
and one attains the Class I\textsubscript{rapid,B} single-wall solution.
This latter solution is compact and entirely distinct from any of the four-dimensional 
phases. After compactification the only access to the conformal boundary is at the 
two points in the far future and past of the Rindler worldline; one may say that the 
dimensionality of the (previously two-dimensional) boundary theory drops to that 
of an instantaneous point. A similar compactification occurs for the Class 
I\textsubscript{rapid,A} and I\textsubscript{rapid,B} single-strut solutions,
although in this case the accessible region of conformal boundary is still 
two-dimensional. The interpretation of such novel compact phases is not clear.

It should also be acknowledged the absence of propagating degrees of 
freedom in three-dimensional gravity creates further distinctions from the 
four-dimensional theory. Consider again the Class I\textsubscript{rapid,A} 
single-wall solution. As we noted, this is similar to a four-dimensional 
accelerating black hole with vanishingly small mass parameter.
However, analytic extension of the four-dimensional solution
necessitates the existence of a ``mirror black hole'' on the other side 
of the acceleration horizon \cite{Dias:2002mi}. For our three-dimensional solution, 
the domain wall was inserted topologically by choosing some value of $x$ at 
which to cut and identify. As such, there is no requirement that one identifies 
similarly in the ``mirror'' region when performing an analytic extension;
one may simply find oneself in a region of pure AdS$_3$ spacetime after 
crossing the event horizon.

We also constructed black holes with domain walls in section \ref{sec:accBTZ}.
In particular, the Class II\textsubscript{left} and Class II\textsubscript{right} 
single-defect solutions each form a one-parameter extension of the standard 
family of static BTZ black holes. We showed how these solutions may be 
constructed from identifications of the Rindler wedge of AdS$_3$,
as the BTZ black hole was by Ba\~{n}ados et al. \cite{Banados:1992gq}.
It is the choice to identify across a surface of constant tension which induces 
the domain wall; the particular choice of identification surface giving zero 
tension produces the static BTZ solution. Apart from the presence of a wall-like 
defect, the Class II\textsubscript{left} solution is qualitatively similar to the usual 
BTZ black hole without a domain wall. It does not form a second horizon for any 
value of the acceleration parameter. However, this is not true of the Class 
II\textsubscript{right} single-strut solution. Here, both ``rapid'' and ``slow'' phases exist.
The rapid phase attained for large acceleration parameter is qualitatively distinct 
from the usual BTZ black hole, possessing a non-compact horizon reminiscent 
of the acceleration horizon which can be present in the four-dimensional AdS 
C-metric geometry. In some sense the solution is a hybrid between the BTZ 
black hole and the four-dimensional AdS C-metric.

However, one should be careful not to carry this analogy too far.
While the conical deficit solutions of section \ref{sec:accptcle} 
may uncontroversially be called accelerating objects, the Class II
BTZ black holes are more subtle. For the particle in three dimensions,
as with the black hole in four dimensions, we can take the mass of the
particle / black hole to zero \emph{at fixed} $A$, and we obtain a Rindler
spacetime -- either a Rindler wedge for rapid acceleration, or a slow
acceleration Rindler coordinatisation of global AdS$_3$.
For these Class II solutions however, taking $m\to0$ at fixed $A$ is not
a meaningful limit -- we cannot remove the black
hole without removing the spacetime altogether! 
For either of the Class II black holes, shrinking the horizon size to zero
requires $m\rightarrow 0$, which in turn requires $K\to\infty$ or 
$x_+\to1$, i.e.\ removing the whole of the spacetime.
Physically, there is then no object with which to ``accelerate'' anything.
Mathematically, in $2+1$ dimensions, Einstein gravity corresponds to a 
topological theory. While the four-dimensional solution is a ``true'' black hole,
the three dimensional black hole is constructed by imposing identifications on the 
acceleration horizon of a fictitious observer \cite{Banados:1992gq}. There is then 
no meaningful sense hitherto established in which one may talk of the acceleration 
of this extended object.
For the Class II\textsubscript{right} black hole in its rapid phase,
the temperature of the non-compact ``acceleration'' horizon also suggests 
that the black hole is inertial.
By regularity of the Euclidean section, the temperature,
with the correct normalisation of Killing vector \eqref{eq:IIrightnormalisation},
is found to be $T = m(2\pi\ell)^{-1}$, which is independent of $\mathcal{A}$.
If the parameter $\mathcal{A}$ did in some way parametrise an acceleration of the black hole,
the Unruh effect would be violated.

An obvious next step in understanding the three-dimensional C-metric
is to establish a consistent thermodynamic description of the system.
Given the four-dimensional results obtained in \cite{Appels:2017xoe,Appels:2016uha,
Anabalon:2018ydc,Anabalon:2018qfv,Gregory:2019dtq,Cassani:2021dwa,
Krtous:2019fpo,Gregory:2020mmi}, one should expect that the tension of the 
domain wall should take the role of a thermodynamic charge. This may also 
provide interesting perspectives from which to understand the role of the defect 
in the boundary theory, since two dimensional conformal field theories are radically 
different from higher-dimensional ones. Realising these topological solutions as 
truncations of supergravity solutions is also an open problem.

Finally, an interesting extension of this work would be to explore the 
quantum accelerating black hole. The dressing of a BTZ black hole
by the back-reaction of a conformal scalar gives rise to $1/r$ corrections
to the geometry \cite{Martinez:1996uv,Casals:2016odj}, but has particular 
interest due to the holographic interpretation as a ``braneworld black hole'' 
\cite{Emparan:1999wa}.
A braneworld is simply a domain wall embedded in a higher dimensional 
AdS spacetime \cite{Randall:1999vf,Randall:1999ee}, but the problem of the 
gravitational field of a black hole on
the brane was never satisfactorily resolved. 
An intriguing analogy however was constructed by Emparan et al.\ by
dropping down a dimension: the four-dimensional C-metric gave a three
dimensional brane with a quantum corrected black hole 
\cite{Emparan:1999wa,Emparan:2020znc}.
Constructing a four dimensional braneworld black hole would therefore give
insight into a four-dimensional quantum corrected black hole. Unfortunately 
however, the five dimensional C-metric proved to be elusive! 
Here, presumably the dressed, accelerating black hole would have similar
$1/r$ corrections to the geometry, however, it is not clear what would be the
four-dimensional holographic counterpart that this 3-brane would slice. 
The temptation is to take a different slicing of the AdS C-metric, however,
the Emparan-Horowitz-Myers slicing of the C-metric is unique 
\cite{Kudoh:2004ub}. We leave this interesting problem for future
investigation.

\begin{acknowledgments}

We would like to thank Pavel Krtou\v{s} for collaboration in the early stages
of this project. We also thank to Aristos Donos and Nabil Iqbal for helpful conversations,
and our anonymous referee for an interesting question on quantum backreaction. 
RG is supported in part by STFC (Consolidated Grant ST/P000371/1),
and by the Perimeter Institute for Theoretical Physics.
Research at Perimeter Institute is supported by the Government of
Canada through the Department of Innovation, Science and Economic 
Development Canada and by the Province of Ontario through the
Ministry of Research, Innovation and Science.
RG would also like to thank the hospitality of the Theoretical Physics Institute at the 
University of Alberta, Edmonton, where this work was initiated.
The work of GAH is funded by Becas Chile (ANID) Scholarship No. 72200271.

\end{acknowledgments}

\appendix

\section{Embedding coordinates for the three-dimensional C-metric}%
\label{app:map}

In this appendix,
we collate mappings from the three-dimensional C-metric geometries
to subsets of global AdS$_3$.
These mappings were used to create the 3D embedding diagrams
in the main text.

\subsection{Global AdS$_3$}
First, we define global AdS$_3$
by its embedding as a hyperboloid in $\mathbb{R}^{2,2}$.
The embedding is
\begin{equation}
\begin{aligned}
  X_0 &=  \ell\sqrt{1+\frac{R^2}{\ell^2}}\sin\left(\frac{T}{\ell}\right)  \,,
  &X_1 &=  R\sin\Theta                                                     \,,
  \\
  X_3 &=  \ell\sqrt{1+\frac{R^2}{\ell^2}}\cos\left(\frac{T}{\ell}\right)  \,,
  &X_2 &=  R\cos\Theta                                                     \,.
\end{aligned}
  \label{eq:globalembedding}
\end{equation}
The hyperboloid is
\begin{equation}
  \sum_{i=0}^{3}\eta_{ii} X_i^2 = \ell^2
  \,, 
\end{equation}
with induced metric
\begin{equation}
  \sum_{i=0}^{3}\eta_{ii} dX_i^2 =
  \left(1+\frac{R^2}{\ell^2}\right)dT^2
  -\frac{dR^2}{\left(1+\frac{R^2}{\ell^2}\right)}
  -R^2d\Theta^2
  \,.
\end{equation}
Note in the above two sums that $\eta$ has the signature $(+--+)$.
The coordinates lie in the ranges
$R\in(0,\infty)$,
$\Theta\in(-\pi,\pi)$,
and $T$ can be taken in the full range of $\mathbb{R}$, with a primary domain of
$T\in [-\pi\ell/2,3\pi\ell/2]$ (chosen for later convenience).

Conversely, the global coordinates are defined from the embedding coordinates by
\begin{equation}
\begin{aligned}
    T_G &= \arctan\left(\frac{X_0}{X_3}\right)  \,,
  & R_G &= \sqrt{X_1^2 +X_2^2}                  \,,\\
    X_G &= X_1                                  \,,
  & Y_G &= X_2                                  \,.
\end{aligned}
  \label{eq:themap}
\end{equation}
We may then compactify the spatial two-section
to attain the Poincar\'{e} disk:
\begin{equation}
  \hat{X} = \frac{X_G}{ \ell + \sqrt{\ell^2 + X_G^2 + Y_G^2} }  \,,\qquad
  \hat{Y} = \frac{Y_G}{ \ell + \sqrt{\ell^2 + X_G^2 + Y_G^2} }  \,.
  \label{eq:thecompactification}
\end{equation}
The global space then appears as a cylinder
in $(T_G,\hat{X},\hat{Y})$ coordinates.
This cylinder is defined by $\left\vert T_G \right\vert < \pi/2$,
$\hat{X}^2+\hat{Y}^2<1$.
Note that this covers only half of the embedding hyperboloid, as
for $T\in [-\pi\ell/2,\pi\ell/2]$ we have $X_3>0$.

Given some metric $ds_3^2$,
by finding a set of embedding coordinates $X_i$
such that
\begin{equation}
  \sum_{i=0}^{3}\eta_{ii} X_i^2 = \ell^2
  \,, 
\end{equation}
and
\begin{equation}
  \sum_{i=0}^{3}\eta_{ii} dX_i^2 = ds_3^2
  \,,
\end{equation}
one may plot the geometry as a subset of the global cylinder
by applying the transformations
\eqref{eq:themap} and \eqref{eq:thecompactification}.

\subsection{The Rindler wedge and the static BTZ black hole}
\label{app:staticBTZ}
For the non-rotating BTZ geometry,
\begin{equation}
  ds^2_3 = 
  \left(-m^2+\frac{r^2}{\ell^2}\right)dt^2
  - \frac{dr^2}{\left(-m^2+\frac{r^2}{\ell^2}\right)}
  - r^2d\phi^2
  \,,
\end{equation}
the embedding is well known \cite{Banados:1992gq}:
\begin{equation}
\begin{aligned}
  X_0 &=  \mathcal{B}(r)\sinh\left(\frac{r_h t}{\ell^2}\right)  \,,
  &X_1 &=  \mathcal{A}(r)\sinh\left(\frac{r_h\phi}{\ell}\right)  \,,\\
  X_3 &=  \mathcal{A}(r)\cosh\left(\frac{r_h\phi}{\ell}\right)  \,,
  &X_2 &= \mathcal{B}(r)\cosh\left(\frac{r_h t}{\ell^2}\right)  \,,
\end{aligned}
\end{equation}
where
\begin{equation}
  \mathcal{A}(r) = \ell\frac{r}{r_h}                         \,,\qquad
  \mathcal{B}(r) = \ell\sqrt{\left(\frac{r}{r_h}\right)^2-1}  \,,
\end{equation}
with
\begin{equation}
  r_h = m\ell \,.
\end{equation}

\begin{figure}[h]
\centering
  \centering
  \includegraphics[width=0.45\linewidth]{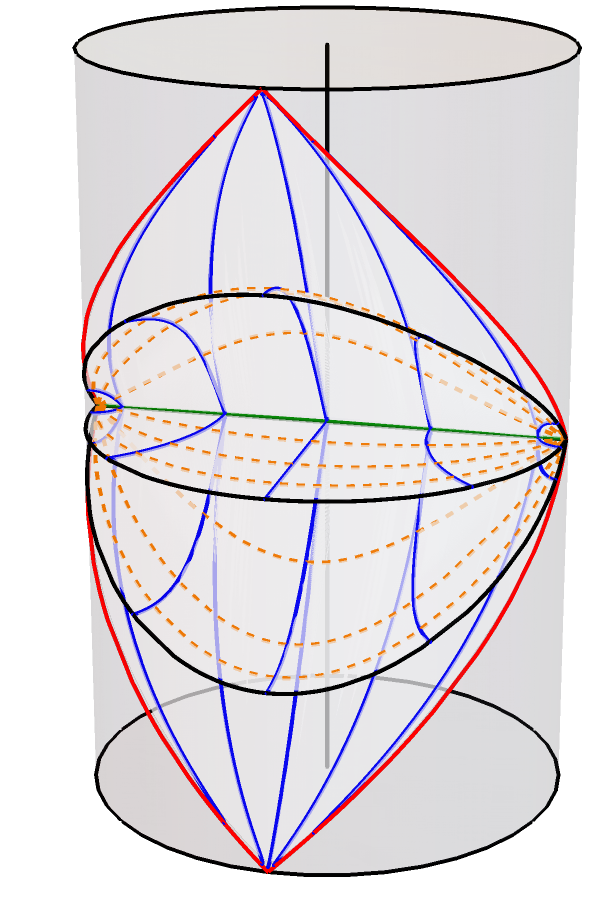}%
  \hfill
  \includegraphics[width=0.45\linewidth]{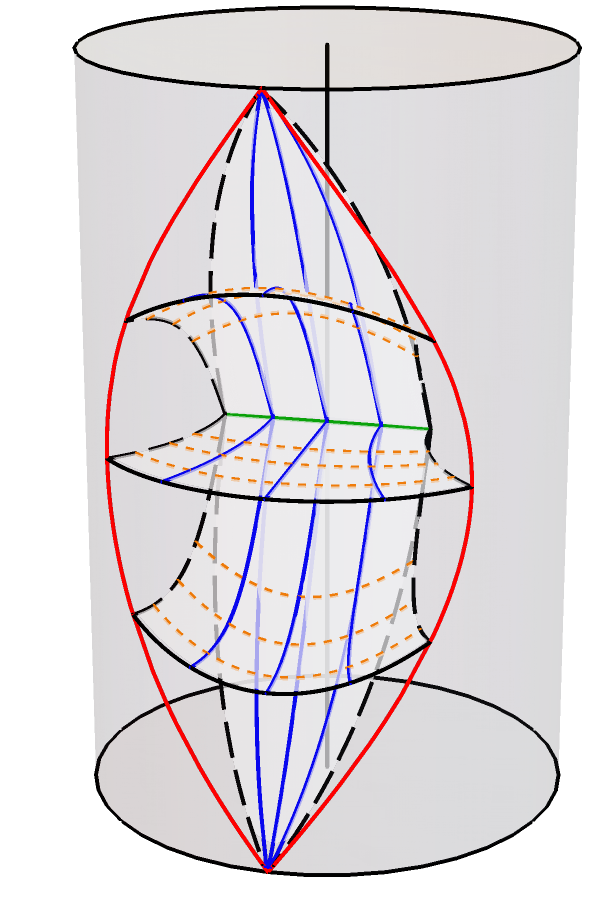}
  \caption{
          The planar and compact (static) BTZ black holes.
          Lines of constant $\phi$ are shown in blue, with lines of constant $r$ in dashed orange.
          Several surfaces of constant time $t$ are shown.
          The classically accesible subset of the boundary is bounded in red.
          To guide the eye, the locus of the cylinder is shown in solid black.
          (a): $\phi\in\mathbb{R}$, giving the planar BTZ geometry or ``Rindler wedge''.
          The bifurcation surface (green) has topology $\mathbb{R}$.
          (b): $\phi$ identified with period $2\pi$, giving the BTZ black hole.
          Lines of constant $t$ and $\phi=\pm\pi$ are plotted in long-dashed black,
          pairs of which are identified across the same time-slice.
          The bifurcation surface (green) has topology $S^1$.
      }
  \label{fig:3Dbtz}
\end{figure}

If $\phi$ is taken to be non-compact,
then we have a portion of AdS$_3$
bounded by a bifurcate acceleration horizon and the conformal boundary
\cite{Banados:1992gq}.
This geometry is the \textit{planar BTZ geometry} or \textit{Rindler wedge}.
In this case $m$ is a gauge parameter which may be set to unity by
a rescaling of both $r$ and $\phi$.
Alternatively, one may identify $\phi\rightarrow\phi+2\pi$
to attain the static BTZ black hole \cite{Banados:1992wn}.
We plot both geometries as subsets of global AdS$_3$
using the technique described above in figure \ref{fig:3Dbtz}.

\subsection{Class I solutions}

The Class I geometry is given by
\begin{equation}
  ds^2_3 = 
  \frac{1}{\Omega^2}\left[Pd\tau^2 - \frac{dy^2}{P} - \frac{dx^2}{Q}\right]
  \,,
\end{equation}
where
\begin{equation}
  Q = 1 - x^2       \,,\qquad
  \Omega = A(x-y)   \,.
\end{equation}

In the slowly accelerating phase, $A^2\ell^2<1$
and the lapse function is given by
\begin{equation}
  P = y^2 + S^2                    \,,\qquad
  S = \sqrt{\frac{1}{A^2\ell^2}-1} \,.
\end{equation}
The embedding is then
\begin{equation}
\begin{aligned}
  X_0 &= \frac{\sqrt{P}}{S\Omega}\sin{S\tau}                \,,
  &X_1 &= \frac{\sqrt{Q}}{\Omega}                            \,,\\
  X_3 &= \frac{\sqrt{P}}{S\Omega}\cos{S\tau}                \,,
  &X_2 &= \frac{A\ell}{\Omega}\left( Sx + \frac{y}{S}\right) \,.
\end{aligned}
\end{equation}

In the rapid phase, $A^2\ell^2> 1$,
so the lapse function now has roots:
\begin{equation}
  P = y^2 - y_h^2                      \,,\qquad
  y_h = \sqrt{1-\frac{1}{A^2\ell^2}}   \,.
\end{equation}
We thus require the alternative embedding
\begin{equation}
\begin{aligned}
  X_0 &=  \frac{\sqrt{P}}{y_h\Omega}\sinh{y_h \tau}                 \,,
  &X_1 &=  \frac{\sqrt{Q}}{\Omega}                                  \,,\\
  X_3 &=  \frac{A\ell}{\Omega}\left| y_h x - \frac{y}{y_h}\right|   \,,
  &X_2 &= \frac{\sqrt{P}}{y_h\Omega}\cosh{y_h \tau}                \,.
\end{aligned}
\end{equation}

\subsection{Class II solutions}

The Class II geometry is given by
\begin{equation}
  ds^2_3 = 
  \frac{1}{\Omega^2}\left[Pd\tau^2 - \frac{dy^2}{P} - \frac{dx^2}{Q}\right]
  \,,
\end{equation}
with metric functions
\begin{equation}
  P = -y^2 + y_h^2                   \,,\qquad
  Q = x^2 - 1                         \,,\qquad
  \Omega = A(x-y)                     \,,
\end{equation}
where
\begin{equation}
  y_h = \sqrt{1+\frac{1}{A^2\ell^2}} \,.
\end{equation}

The embedding in the region where $P$ is positive is then
\begin{equation}
\begin{aligned}
  X_0 &=  \frac{\sqrt{P}}{y_h\Omega}\sinh{y_h \tau}               \,,
  &X_1 &= \frac{\sqrt{Q}}{\Omega}                                 \,,\\
  X_3 &=  \frac{A\ell}{\Omega}\left( y_h x - \frac{y}{y_h}\right) \,,
  &X_2 &=  \frac{\sqrt{P}}{y_h\Omega}\cosh{y_h \tau}              \,.
\end{aligned}
\end{equation}

\subsection{Class III solutions}
For the Class III solution of section \ref{sec:III},
the line element is
\begin{equation}
  ds^2_3 = 
  \frac{1}{\Omega^2}\left[Pd\tau^2 - \frac{dy^2}{P} - \frac{dx^2}{Q}\right]
  \,.
\end{equation}
The metric functions are
\begin{equation}
  P = -y^2 + y_h^2                             \,,\qquad
  Q = 1 + x^2                                  \,,\qquad
  \Omega = A(x-y)                              \,,
\end{equation}
where
\begin{equation}
  y_h = \sqrt{\frac{1}{A^2\ell^2}-1}           \,.
\end{equation}
The embedding for $\left\vert y \right\vert < y_h$ is then
\begin{equation}
\begin{aligned}
  X_0 &=  \frac{\sqrt{P}}{y_h\Omega}\sinh{y_h \tau}                 \,,
  &X_1 &= \frac{A\ell}{\Omega}\left( y_h x + \frac{y}{y_h}\right)   \,,\\
  X_3 &=  \frac{\sqrt{Q}}{\Omega}                                   \,,
  &X_2 &= \frac{\sqrt{P}}{y_h\Omega}\cosh{y_h \tau}                 \,.
\end{aligned}
\end{equation}

\end{document}